\documentclass[prl, twocolumn, superscriptaddress]{revtex4-1}
\usepackage[version=3]{mhchem} 

\usepackage{xr}
\usepackage[english]{babel}
\usepackage{natbib}
\usepackage{microtype}
\usepackage{hyperref}
\usepackage{MnSymbol}%
\usepackage{wasysym}%
\usepackage{graphicx}

\hypersetup{colorlinks=true,citecolor=blue}
\graphicspath{{./images/}}

\newcommand{\Delft}{Kavli Institute of Nanoscience, Delft University of Technology, 2600 GA Delft, The Netherlands.}
\newcommand{\Konstanz}{Department of Physics, University of Konstanz, D-78464 Konstanz, Germany.}
\newcommand{\NIMS}{National Institute for Materials Science, 1-1 Namiki, Tsukuba, 305-0044, Japan.}
\newcommand{\Tubingen}{Physikalisches Institut and Center for Quantum Science (CQ) in LISA$^+$, Eberhard Karls Universit\"{a}t T\"{u}bingen, Auf der Morgenstelle 14, D-72076 T\"{u}bingen, Germany.}
\newcommand{\Budapest}{Department of Physics of Complex Systems, E\"{o}tv\"{o}s University, P\'{a}zm\'{a}ny P\'{e}ter S\'{e}t\'{a}ny 1/A, H-1117 Budapest, Hungary.}
\newcommand{\Austria}{Institute of Science and Technology Austria, Am Campus 1, A-3400 Klosterneuburg, Austria.}
\newcommand{\Qutech}{QuTech, Delft University of Technology, 2600 GA Delft, The Netherlands.}

\newcommand{\TITLE}{Current-phase relation of ballistic graphene Josephson junctions}

\begin{document}

\title{\TITLE}
\author{G.~Nanda}\affiliation{\Delft}
\author{J.~L.~Aguilera-Servin}\affiliation{\Delft}\affiliation{\Austria}
\author{P.~Rakyta}\affiliation{\Budapest}
\author{A.~Korm\'{a}nyos}\affiliation{\Konstanz}
\author{R.~Kleiner}\affiliation{\Tubingen}
\author{D.~Koelle}\affiliation{\Tubingen}
\author{K.~Watanabe}\affiliation{\NIMS}
\author{T.~Taniguchi}\affiliation{\NIMS}
\author{L.~M.~K.~Vandersypen}\affiliation{\Delft}\affiliation{\Qutech}
\author{S.~Goswami}\email{S.Goswami@tudelft.nl}\affiliation{\Delft}\affiliation{\Qutech}

\begin{abstract}

The current-phase relation (CPR) of a Josephson junction (JJ) determines how the supercurrent evolves with the superconducting phase difference across the junction. Knowledge of the CPR is essential in order to understand the response of a JJ to various external parameters. Despite the rising interest in ultra-clean encapsulated graphene JJs, the CPR of such junctions remains unknown. Here, we use a fully gate-tunable graphene superconducting quantum intereference device (SQUID) to determine the CPR of ballistic graphene JJs. Each of the two JJs in the SQUID is made with graphene encapsulated in hexagonal boron nitride. By independently controlling the critical current of the JJs, we can operate the SQUID either in a symmetric or asymmetric configuration. The highly asymmetric SQUID allows us to phase-bias one of the JJs and thereby directly obtain its CPR. The CPR is found to be skewed, deviating significantly from a sinusoidal form. The skewness can be tuned with the gate voltage and oscillates in anti-phase with Fabry-P\'{e}rot resistance oscillations of the ballistic graphene cavity. We compare our experiments with tight-binding calculations which include realistic graphene-superconductor interfaces and find a good qualitative agreement. 
\end{abstract}

\maketitle
\setcounter{table}{0}
\renewcommand{\thetable}{\arabic{table}}%
\setcounter{figure}{0}
\renewcommand{\thefigure}{\arabic{figure}}%

The past few years have seen remarkable progress in the study of graphene-superconductor hybrids. This surge in interest has primarily been driven by the ability to combine high-quality graphene with superconductors via clean interfaces, and has led to several experimental breakthroughs. These include the observation of specular Andreev reflection~\cite{Efetov}, crossed Andreev reflections~\cite{Kim2}, and superconducting proximity effects in ballistic graphene Josephson junctions (JJs)~\cite{Calado,Shalom,Allen,Amet,Borzenets}. In a majority of these studies the device comprises of graphene encapsulated in hexagonal boron nitride (BN) contacted along the edge by a superconductor. The encapsulation in BN keeps the graphene clean, while the edge contacting scheme provides transparent interfaces. In particular, ballistic JJs fabricated in this manner have been central to recent studies of novel Andreev bound states in perpendicular magnetic fields~\cite{Shalom}, edge-mode superconductivity~\cite{Allen}, and supercurrents in the quantum Hall regime~\cite{Amet}. However, to date there have been no measurements of the Josephson current phase relation (CPR) in these systems.

The CPR is arguably one of the most basic properties of a JJ, and provides information about the Andreev bound state (ABS) spectrum in the junction. While typical superconductor-insulator-superconductor (SIS) JJs exhibit a sinusoidal CPR, deviations from this behavior can be present in superconductor-normal-superconductor (SNS) junctions. Examples of this include JJs with high transmission such as nanowires~\cite{Murani_Bi} and atomic point contacts~\cite{Ootuka,Rocca},  where the CPR contains significant higher frequency components. Furthermore, the periodicity of the CPR itself can be different from $2\pi$ for more exotic systems such as topological JJs~\cite{Molemkamp_4pi}. For graphene JJs there have been several numerical estimates of the CPR which take into account its linear dispersion relation~\cite{Titov,Cserti,Annica_1,Linder,Rakyta}. More recently, ballistic graphene JJs operated in large magnetic fields have been predicted to undergo a topological transition~\cite{Pablo_Majo} which should be detectable via direct CPR measurements. However, the experimental determination of the CPR in graphene has been restricted to junctions which are either in the diffusive limit~\cite{Harlingen} or in a geometry which does not allow gate control of the junction properties~\cite{Lee}. 

 \begin{figure*}[!t]
 	\centering
 	\includegraphics[width=1\linewidth]{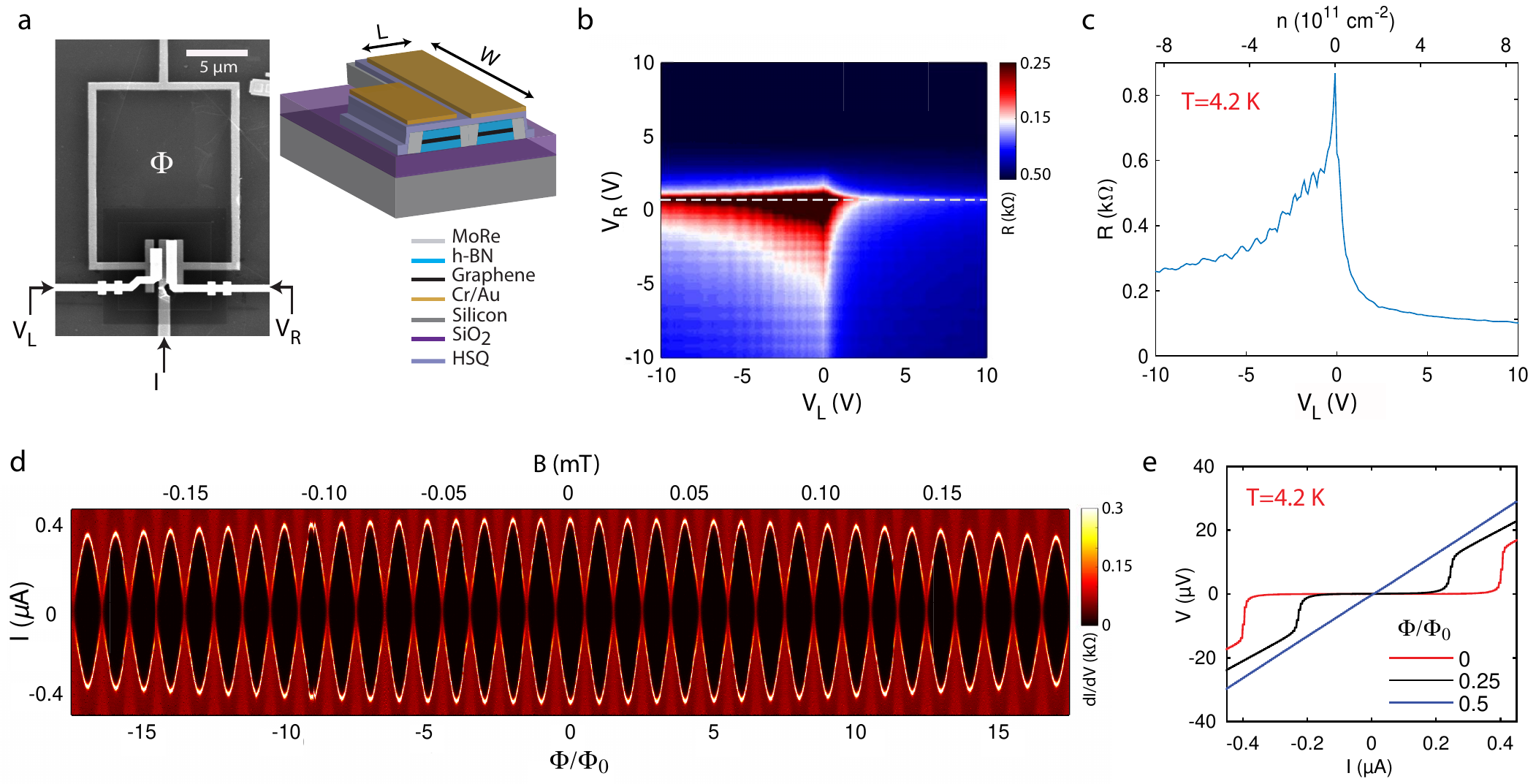}%
 	\caption{(a) Scanning electron micrograph of the graphene dc-SQUID (Dev1) along with a cross-sectional schematic. Gate voltages $V_L$ and $V_R$ independently control the carrier density of the left and right junction respectively. (b) Resistance $R$ across the SQUID vs $V_{L}$ and $V_{R}$, demonstrating independent control of carrier type and density in the JJs. (c) Line trace taken along the dashed white line in (b) showing Fabry-P\'{e}rot oscillations in the hole-doped regime. (d) Differential resistance $dV/dI$ as a function of dc current bias $I$ and magnetic field $B$, with the SQUID operated in a symmetric configuration ($V_{L}=+10$~V and $V_{R}=+2.5$~V). Flux-periodic oscillations are clearly visible with a slowly decaying envelope arising from the interference pattern of a single JJ. (e) $V-I$ plots [extracted from (d)] for different values of magnetic flux $\Phi$ showing a nearly 100~\% modulation of the critical current. All measurements shown here are performed at $T = 4.2$~K. \label{fig:SQUID}}
 \end{figure*}
 
Here, we use a dc superconducting quantum interference device (SQUID) to directly determine the CPR in encapsulated graphene JJs. These graphene SQUIDs stand out from previous studies~\cite{Girit,Girit2} in two important ways. Firstly, the superconducting contacts are made with Molybdenum Rhenium (MoRe), which allows us to operate the SQUID up to 4.2~K. More importantly, our SQUID consists of graphene JJs which are ballistic and independently tunable, thereby allowing full electrostatic control over the SQUID response. By applying appropriate gate voltages we can continuously tune from a symmetric to an asymmetric SQUID. We show that the asymmetric configuration allows us to directly extract the CPR from flux periodic oscillations in the critical current of the SQUID. The CPR is found to be non-sinusoidal, displaying a prominent forward skewing. This skewness can be tuned over a large range with the gate voltage and shows correlations with Fabry-P\'{e}rot (FP) resistance oscillations in the ballistic cavity. We complement our experiments with tight-binding simulations which go beyond the short junction limit and explicitly take into account realistic graphene-superconductor interfaces.

Figure~\ref{fig:SQUID}a shows a scanning electron micrograph and cross-sectional schematic of a device. It consists of two encapsulated graphene JJs contacted with MoRe, incorporated in a SQUID loop. The fabrication strategy is similar to earlier work~\cite{Calado} and further details are provided in the Supplementary Information (SI). The left~(L-JJ)/right~(R-JJ) JJs can be tuned independently by applying voltages ($V_L$/$V_R$) to local top gates. The junctions are intentionally designed to have a geometrical asymmetry, which ensures that the critical current of R-JJ~($I_{cR}$) is larger than that of L-JJ~($I_{cL}$) at the same carrier density. We report on two devices (Dev1 and Dev2) both of which have the same lithographic dimensions (L~$\times$~W) for L-JJ (400~nm~$\times$~2~$\mu$m). The dimensions of R-JJ for Dev1 and Dev2 are 400~nm~$\times$~4~$\mu$m and 400~nm~$\times$~8~$\mu$m respectively. All measurements were performed using a dc current bias applied across the SQUID, in a dilution refrigerator with a base temperature of 40~mK.

Figure~\ref{fig:SQUID}b shows the variation in the normal state resistance ($R$) of the SQUID with $V_L$ and $V_R$ at $T = 4.2$~K. The device was biased with a relatively large current of 500~nA, which is larger than the critical current of the SQUID for most of the gate range. Figure~\ref{fig:SQUID}c shows a single trace taken along the white dashed line of Figure~\ref{fig:SQUID}b, where R-JJ is held at the charge neutrality point (CNP). We see clear FP oscillations on the hole ($p$) doped region due to the formation of $n-p$ junctions at the superconductor-graphene interfaces~\cite{Calado,Shalom}. Furthermore, the criss-cross pattern seen in the lower left quadrant of Figure~\ref{fig:SQUID}b indicates that both graphene junctions are in the ballistic limit and that there is no cross-talk between the individual gates. We note that when $V_R > 3$~V the critical current of the SQUID ($I_c$) is larger than the applied current bias, and a zero-resistance state is thus visible even at 4.2~K. Having established the fact that our JJs are in the ballistic regime, we now look in more detail at the behavior of the SQUID. At $T=4.2$~K we first tune the gate voltages ($V_L=+10$~V, $V_R=+2.5$~V) such that the SQUID is in a symmetric configuration and $I_{cR} = I_{cL}$. Figure~\ref{fig:SQUID}d shows the variation in differential resistance $dV/dI$ with current bias $I$ and magnetic field $B$, where we observe clear oscillations in $I_c$ with magnetic flux. In this configuration, the modulation in $I_c$ is nearly 100~\%, as seen by the individual $V-I$ traces in Figure~\ref{fig:SQUID}e. The slow decay in the maximum value of $I_c$ arises due to the (Fraunhofer) magnetic field response of a single junction. The devices were designed such that this background was negligible around $B=0$ (i.e., the SQUID loop area was kept much larger than the JJ area). Minimizing this background is important for a reliable determination of the CPR, as we will see below.

\begin{figure}[!t]
	\centering
	\includegraphics[width=1\linewidth]{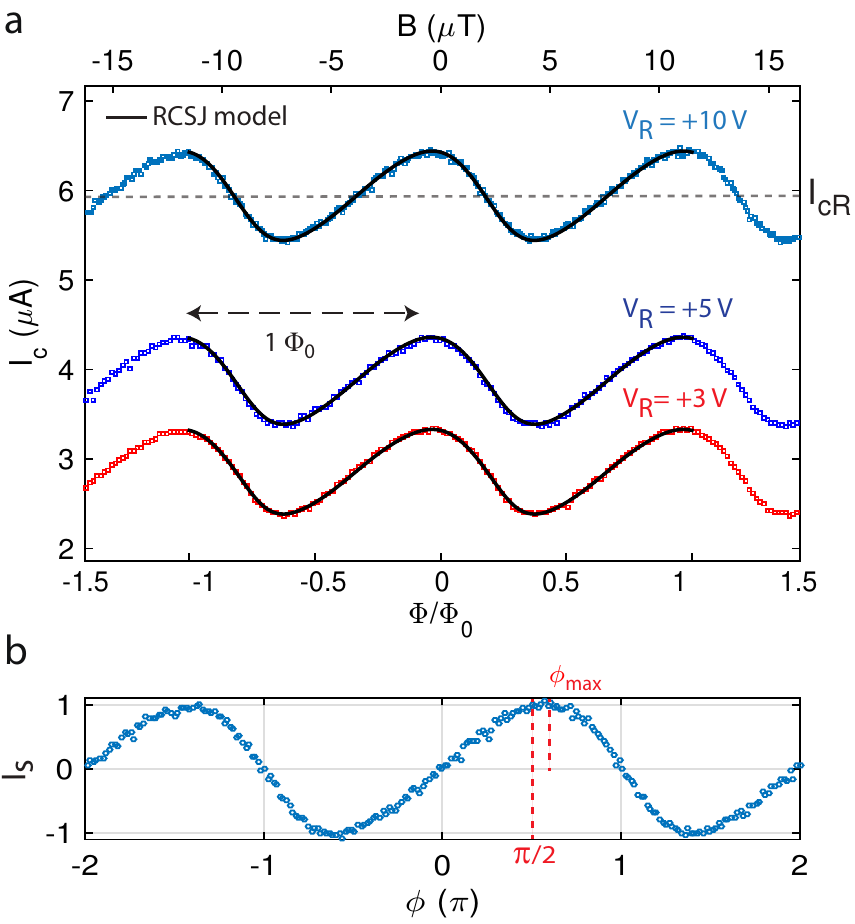}%
	\caption{(a) Variation of $I_{c}$ with $\Phi$ for $V_{L} =-4$~V and $V_{R}=+10$~V, $+5$~V and $+3$~V at 40~mK. Solid black lines are results from RCSJ simulations of the SQUID. (b) Variation of supercurrent $I_s = (I_c-I_{cR})/I_{cL}$ with phase $\phi$ extracted from the top curve in (a). $\phi_{max}$ indicates the phase at which $I_s$ reaches a maximum, and is noticeably different from $\pi/2$, indicating a forward skewed CPR.\label{fig:CPR}}
\end{figure}

We now turn our attention to the flux-dependent response of a highly asymmetric SQUID ($I_{cR}>>I_{cL}$), a condition which can be readily achieved by tuning the gate voltages appropriately. To outline the working principle of the device, we start with the assumption that both JJs have a sinusoidal CPR (a more general treatment can be found in the SI). So, the critical current of the SQUID can be written as $I_{c}=I_{cL}\sin\theta + I_{cR}\sin\delta$, where $\theta$ ($\delta$) is the phase drop across L-JJ (R-JJ). When an external magnetic flux ($\Phi$) threads through the SQUID loop, the flux and phase are related by $\delta-\theta= 2\pi\Phi/\Phi_{0}$, assuming the loop inductance is negligible. Now, when $I_{cR}>>I_{cL}$ the phase difference across R-JJ is very close to $\pi/2$. Thus, $I_{c} (\Phi) \approx I_{cR} + I_{cL} \sin(2\pi\Phi/\Phi_{0}+\pi/2)$ and the flux-dependence of $I_{c}$ directly represents the CPR of L-JJ, i.e., $I_{c}(\Phi) \approx I_{cR} + I_s(\phi)$, where $I_s$ is the supercurrent through L-JJ and $\phi$ is the phase drop across it. This principle of using an asymmetric SQUID to probe the CPR has been employed in the past for systems such as point contacts~\cite{Ootuka,Rocca} and vertical graphene JJs~\cite{Lee}, where an SIS junction (with a well known sinusoidal CPR) was used as the reference junction. In our case, the reference junction is also a graphene JJ, where the CPR is not known a priori. We show (see SI) that this does not affect our ability to probe the CPR, provided time reversal symmetry is not broken, meaning that the CPR satisfies the condition $I_s(\phi) = -I_s (-\phi)$~\cite{Golubov}. Throughout the remainder of the text we use R-JJ as the reference junction (larger critical current), and L-JJ is the junction under study (smaller critical current).

Figure~\ref{fig:CPR}a shows the typical magnetic response of the asymmetric SQUID at $T=40$~mK, with $V_L = -4$~V (fixed) and different values of $V_R$. For the most asymmetric configuration ($V_R = +10$~V) $I_c$ oscillates around a fixed value of roughly 6~$\mu$A ($I_{cR}$) with an amplitude of about 500~nA ($I_{cL}$). Using the arguments described above, this $I_c (\Phi)$ curve can be converted to $I_s (\phi)$, as shown in Figure~\ref{fig:CPR}b. Here $I_s$ is the normalized supercurrent defined as $(I_c-I_{cR})/I_{cL}$. We note that there is an uncertainty (less than one period) in the exact position of zero $B$. This, combined with the unknown CPR of the reference graphene JJ, makes it important to do this conversion carefully, and we describe the details in the SI. The CPR shows a clear deviation from a sinusoidal form, showing a prominent forward skewing (i.e., $I_s$ peaks at $\phi>\pi/2$). We define the skewness of the CPR as $S= (2\phi_{max}/\pi)-1$~\cite{Harlingen}, where $\phi_{max}$ is the phase for which the supercurrent is maximum. 

To be certain that we are indeed measuring the CPR of L-JJ, we perform some important checks. We keep $I_{cL}$ fixed and reduce $I_{cR}$ (by reducing $V_R$). Figure~\ref{fig:CPR}a shows that reducing $I_{cR}$ merely shifts the $I_c(\Phi)$ downwards and therefore does not affect the extracted CPR, as one would expect. Furthermore, we use the experimentally determined CPR (from Figure~\ref{fig:CPR}b), the junction asymmetry, and loop inductance as inputs for the resistively and capacitively shunted junction (RCSJ) model to compute the expected SQUID response (see SI for details of the simulations). These plots (solid lines) show an excellent agreement between simulations and experiment, thus confirming that the asymmetry of our SQUID is sufficient to reliably estimate the CPR of L-JJ. Furthermore, it shows that there are no significant effects of inductance in our measurements, which could potentially complicate the extraction of the CPR from $I_c (\Phi)$ in an asymmetric SQUID~\cite{Fulton}.

\begin{figure}[!t]
	\includegraphics[width=1\linewidth]{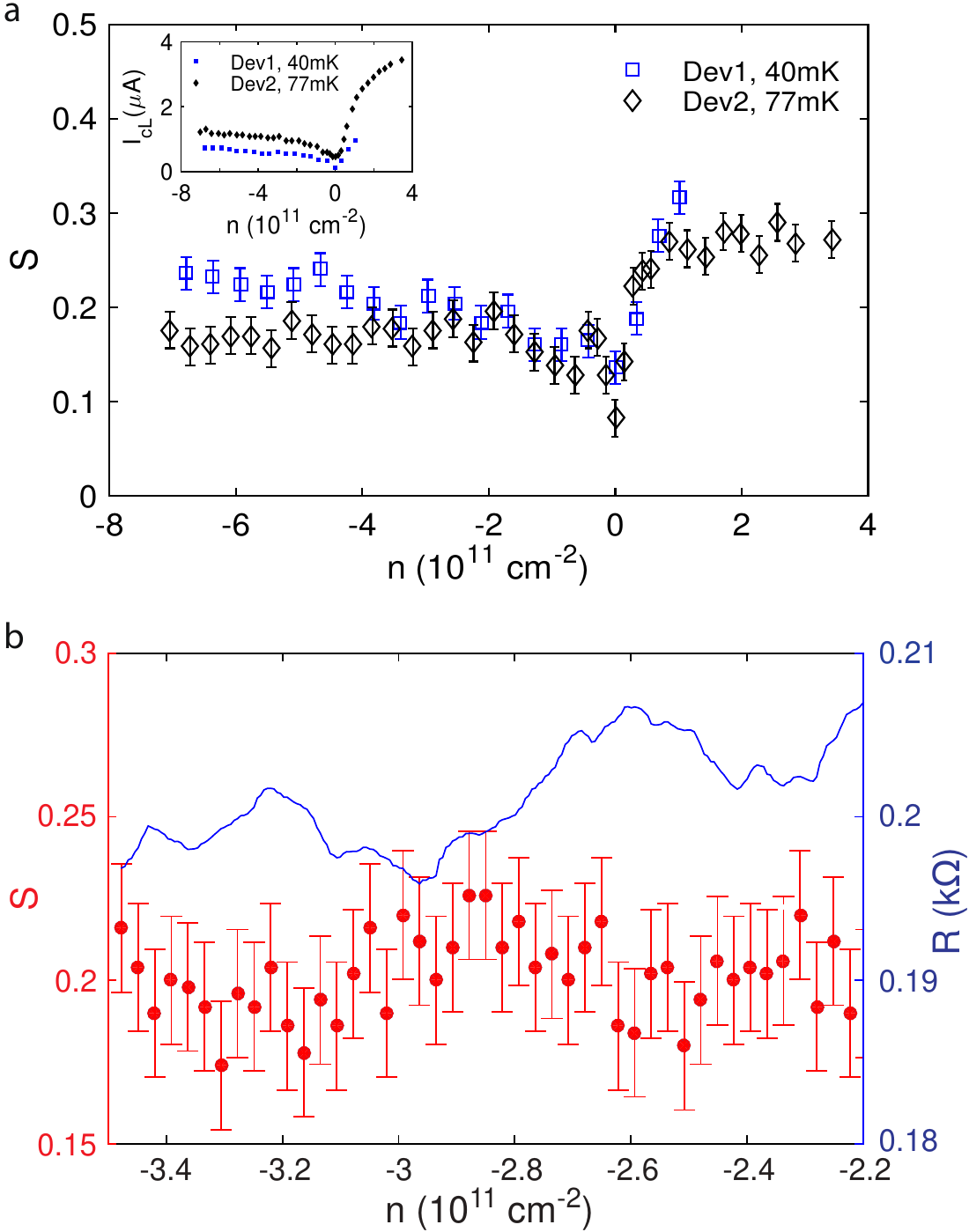}%
	\caption{ (a) Variation of skewness $S$ as a function of carrier density $n$ for Dev1 and Dev2. The larger geometric asymmetry of Dev2 (see text) allows one to reliably probe the CPR up to higher n-doping. Inset shows the variation of $I_{cL}$ with density. (b) $S$ oscillates with carrier density in the p-doped regime in anti-phase with Fabry P\'{e}rot oscillations in the resistance.\label{fig:CPR2}}
\end{figure} 

To study the gate dependence of the CPR we fix $V_R$ at +10~V (to maximize $I_{cR}$) and study the change in $S$ with $V_L$ (Figure~\ref{fig:CPR2}a) for Dev1 and Dev2. For both devices we find that $S$ is larger on the n-side as compared to the p-side, showing a dip close to the CNP. We note that Dev2 allows us to probe the CPR up to a larger range on the n-side due to its larger geometric asymmetry (see SI for other measurements on Dev2). We expect the skewness to depend strongly on the total transmission through the graphene JJ, which should depend on (a) the number of conducting channels in the graphene, as well as (b) the transparency of the graphene-superconductor interface. The gate voltage $V_L$ obviously changes the Fermi wave vector, but it also changes the contact resistance~\cite{Wang}, which plays a significant role in determining $S$. This can be seen most clearly for Dev2 for high n-doping, where $S$ saturates, despite the fact that $I_{cL}$ continues to increase up to the largest measured density (see inset). At large p-doping $S$ also seems to saturate, but a closer look (Figure~\ref{fig:CPR2}b) shows that $S$ oscillates in anti-phase with the FP oscillations in resistance. This clearly indicates that in this regime the CPR is modulated by phase coherent interference effects similar to the FP oscillations which affect the total transmission. 

\begin{figure*}[ht!]
	\centering
	\includegraphics[width=0.75\linewidth]{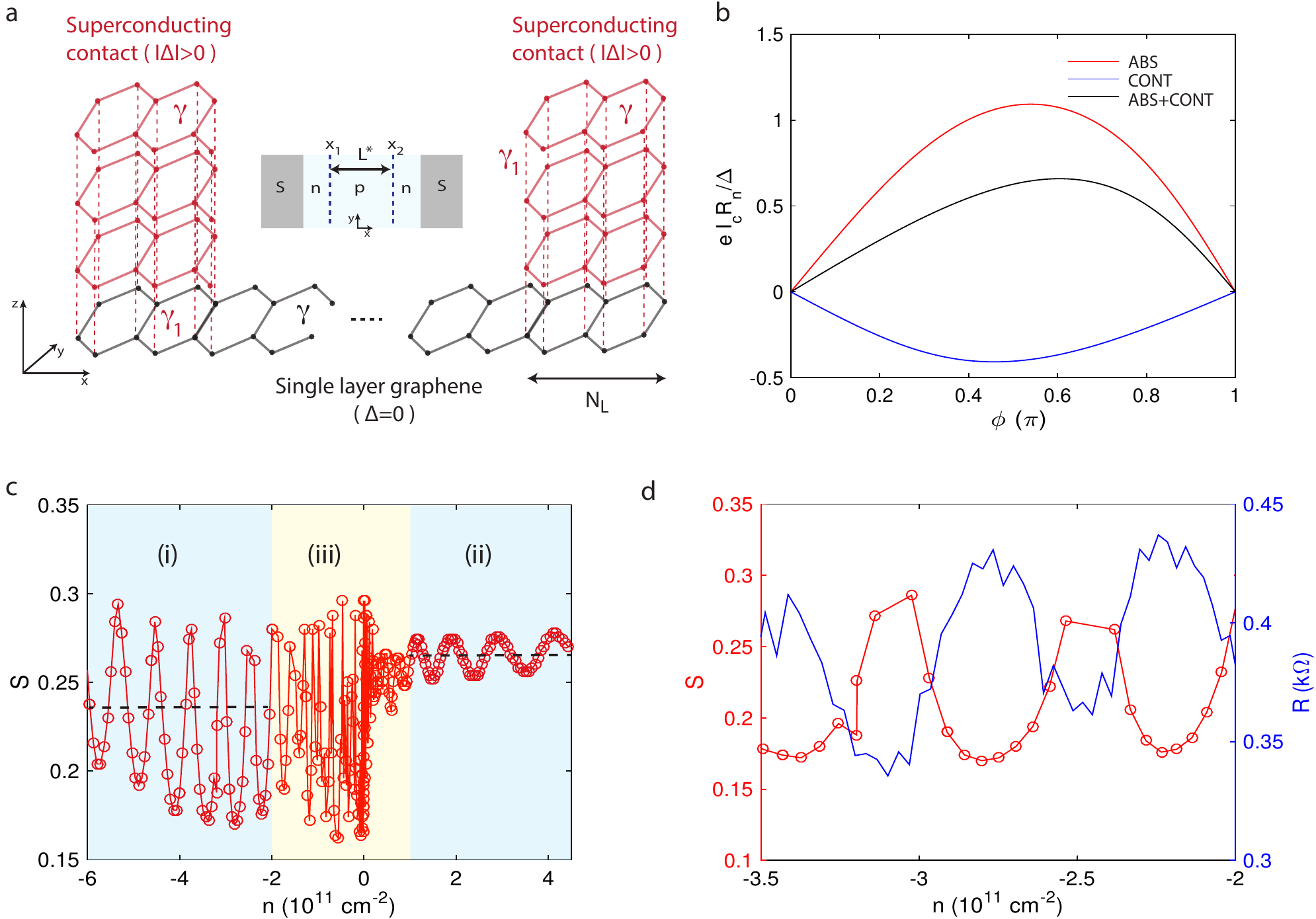}
	\caption{(a) The geometry of the system used in the calculations. The superconducting leads are attached in a top-contact geometry to the normal graphene sheet. A periodic boundary condition is applied in the $y$ direction. (Inset) Top view of the system. Due to doping from the S contacts, the normal graphene region is assumed to be n-doped up to a distance $x_1$ ($x_2$) from the left (right) contact. The distance $L^{*}=x_2-x_1$ is the effective cavity length which depends on the gate voltage applied to the junction. (b) The contribution of the ABSs (red) and continuum CONT (blue) to the total supercurrent (black) as a function of the phase difference for an n-doped junction. (c) The skewness $S$ as a function of doping of the junction.  The regimes i-iii indicated by the rectangles are further discussed in the text. Dashed lines show the  average $S$ in the $p$ and $n$ doped regime. (d) The skewness (red circles, left axis) and  normal state resistance (blue, right axis) vs doping for strong $p$-doping of the junction.\label{fig:THEORY}}
\end{figure*}

We complement our measurements with a minimal theoretical model by solving the corresponding Bogoliubov-de Gennes (BdG) equations to calculate the CPR in graphene JJs. To set the stage, we note that SNS junctions can be characterized by the quasiparticle mean free path $l_{f}$ in the normal (N) region and the coherence  length $\xi_0=\hbar v_F/\Delta$, where $v_F$ is the Fermi velocity in N. In our devices $L\ll l_f$, i.e., they are in the ballistic regime, and therefore we neglect impurity scattering in our calculations. Taking $v_F\approx 10^{6}$~m/s for graphene and  $\Delta\approx 1.2\,$~meV for MoRe, one finds $\xi_0=548$~nm, which means that in our junctions   $L \lesssim \xi_0$, i.e., they are not in the strict short junction limit $L\ll \xi_0$. Consequently, the Josephson current is carried not only by discrete Andreev bound states (ABSs), but also by states in the continuum (CONT)~\cite{bratus,affleck,cini}. To this end we numerically solve the BdG equations using a tight-binding (TB) model (see Figure~\ref{fig:THEORY}a) and a recently developed numerical approach~\cite{Rakyta,Equus} which handles the ABSs and states in the continuum on equal footing. The description of both the normal region and the superconducting terminals is based on the nearest-neighbor TB model of graphene~\cite{graphene-TB}. The on-site complex pair-potential $\Delta$ is finite only in the superconducting terminals and changes as a step-function at the N-S interface. The results presented here are calculated using the top-contact geometry (Figure~\ref{fig:THEORY}a), a model with perfect edge contacts is discussed in the SI. As observed experimentally, we take into account $n$-doping from the superconducting contacts (see Figure~\ref{fig:THEORY}b). If the junction is gated into hole-doping, a FP cavity is formed by the two $n-p$ junctions in the vicinity of the left and right superconducting terminals. The length $L^{*}$  of this FP cavity depends on the gate voltage~\cite{Shalom}, for stronger hole-doping the n-p junctions shift closer to the contacts. For further details of the model see SI.

Turning now to the CPR calculations, Figure~\ref{fig:THEORY}b shows separately the contribution of the ABS and the continuum to the supercurrent. Since $L\lesssim \xi_0$, the latter contribution is not negligible and affects both the value of the critical current and the skewness of the CPR. In Figure~\ref{fig:THEORY}c we show the calculated skewness $S$ as a function of the doping of the junction at zero temperature. We consider three regimes: (i) strongly $p$-doped  junction; (ii) large $n$-doping, (iii) the region around the CNP. We start with the discussion of (i). It is well established that in this case the $p$-$n$ junctions lead to FP oscillations in the normal resistance as well as in the critical current~\cite{Calado,Shalom} of graphene JJs. Our calculations,  shown in Figure~\ref{fig:THEORY}d, indicate that  due to FP interference the skewness also displays oscillations as a function of doping around an average value of $S \approx 0.23$. As already mentioned, similar oscillations are present in the normal state resistance $R$, however, we find that $R$ oscillates  in antiphase with the skewness. Compared to the measurements (Figure~\ref{fig:CPR2}b), our calculations therefore reproduce the phase relation between the oscillations of the skewness and $R$ and give a qualitatively good agreement with the measured values of the skewness. In the strong $n$-doped regime (case ii) the calculated average skewness of $S=0.27$  is larger than for $p$-doped junctions, and very close to the measured values. Small oscillations of $S$ are still present in our results and they are due to the $n$-$n'$ interfaces, i.e., the  difference in the doping close to the contacts (for $x<x_1$ and $x>x_2$) in Figure~\ref{fig:THEORY}a and the junction region ($x_1<x<x_2$), which enhances backscattering. Our calculations therefore predict smaller skewness for $p$-doped than for $n$-doped junctions. The enhancement of $S$ in the $n$-doped regime can be clearly seen in the measurements of Figure~\ref{fig:CPR2}a. We note that previous theoretical work~\cite{Linder}, which calculated the spatial dependence of the pairing amplitude self-consistently, predicted a skewness of $S\approx 0.15$ for $n$-doped samples with $L<\xi_0$, while a non-self-consistent calculation which took into account only the contribution of the ABS yielded $S \approx 0.42$~\cite{Linder}. The comparison of these results to ours, and to the measurements, suggest that the skewness may depend quite sensitively on the S-N interface as well as on the $L/\xi_0$ ratio and that our approach captures the most important effects in these junctions. Finally, we briefly discuss the case (iii), where the measurements show a suppression of the skewness as the CNP is approached. The measured values of $S\sim 0.1$ are similar to those found in diffusive junctions~\cite{Harlingen}, but significantly lower than the theoretical prediction of $S=0.26$ in the short junction limit~\cite{Titov} at the CNP. This suppression of $S$ is not reproduced in our calculations, instead, we find rapid oscillations as the CNP is approached from the $p$-doped regime. This discrepancy is likely to be due to effects that are not included in our ballistic model, such as  charged scatterers which are poorly screened in this regime, or scattering at the edges, which is  more relevant at low densities when only a few open channels are present. 

\begin{figure}[!tb]
	\centering
	\includegraphics[width=1\linewidth]{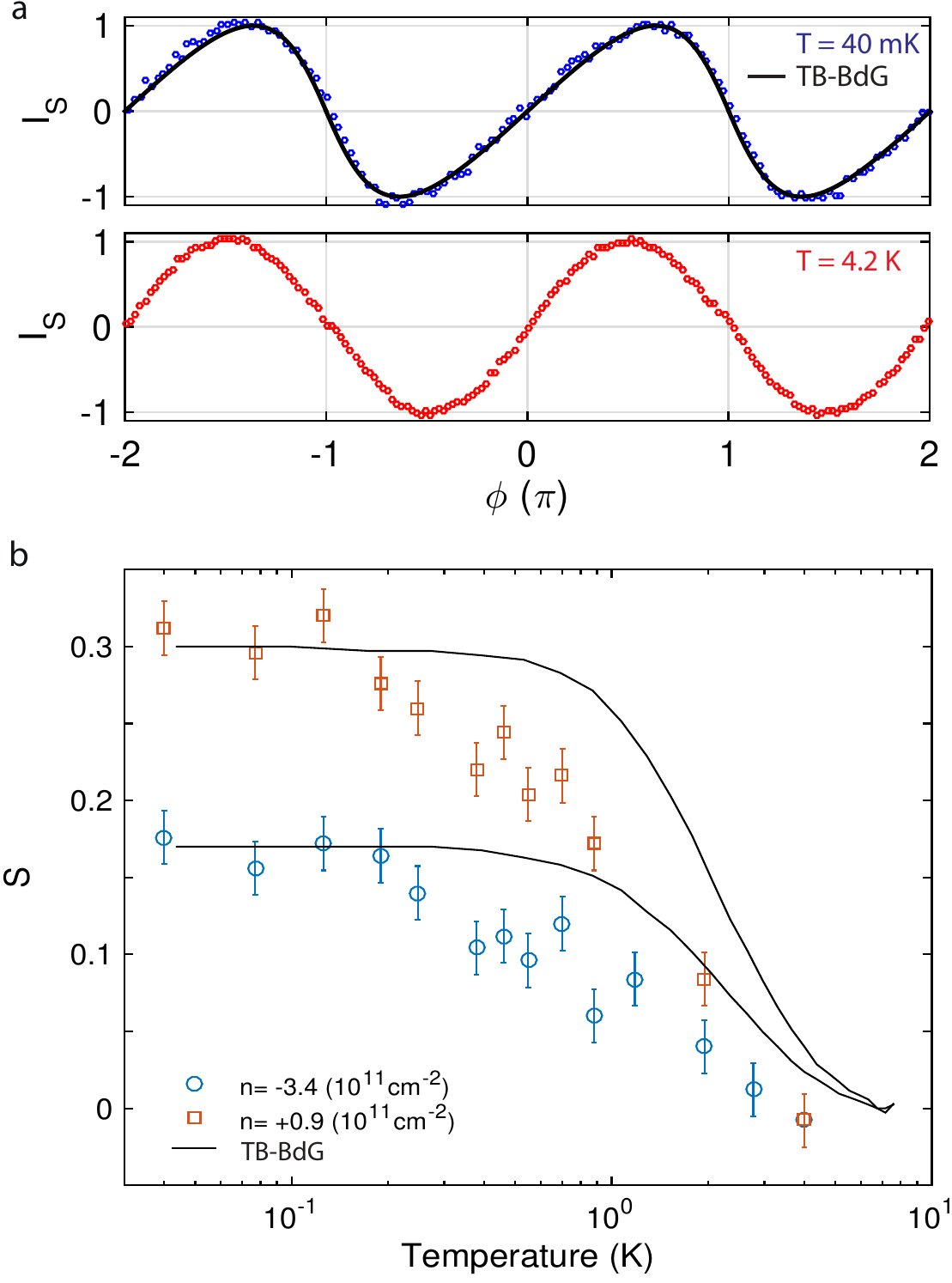}%
	\caption{(a) CPR for $V_{L}$=+1 V ($n = 0.9 \times 10^{11}$~cm$^{-2}$) at 40 mK (upper curve) and 4.2~K (lower curve). Solid line shows the calculated CPR. A forward skewness is clearly seen in the curve at 40~mK but is absent at 4.2~K. (b) Variation of $S$ with temperature for electron and hole doping. Increasing the temperature suppresses higher harmonics in the CPR, thereby reducing $S$ until it vanishes near 4.2~K and the curves become sinusoidal. Black lines show the results of tight binding simulations.}
		\label{fig:TEMP}%
\end{figure}

Finally, we study the effect of temperature on the CPR of these JJs. In Figure~\ref{fig:TEMP}a, we compare the CPR in the n-doped regime ($V_{R} = +1$~V; $n = 0.9 \times 10^{11}$~cm$^{-2}$) at 40~mK and 4.2~K. One clearly sees that at 4.2~K the CPR is sinusoidal. This is consistent with our observation that the critical current modulation of the SQUID is nearly 100~\% at 4.2~K (Figure~\ref{fig:SQUID}d), a condition which can only be achieved if the CPR is sinusoidal. Figure~\ref{fig:TEMP}b shows the full temperature dependence of $S$ for two representative values of electron and hole doping. The reduction in skewness with temperature is a consequence of the fact that the higher frequency terms in the CPR arise due to the phase coherent transfer of multiple Cooper pairs and involve longer quasiparticle paths~\cite{heikkila}, thereby making them more sensitive to temperature. As a result, their amplitude decreases quickly with increasing temperature~\cite{Cserti,Linder,Rakyta,Harlingen}.  Numerical estimates show the same qualitative behavior, however the experimentally determined skewness reaches zero (sinusoidal CPR) faster than the numerics. At this point it is difficult to ascertain the exact reason for this discrepancy, but one possible explanation for this is that the induced superconducting gap in the graphene is somewhat smaller than the bulk MoRe gap, resulting in a faster decay.  

In conclusion, we have used a fully gate-tunable graphene based SQUID to provide measurements of the current-phase relation in ballistic Josephson junctions made with encapsulated graphene. We show that the CPR is non-sinusoidal and can be controlled by a gate voltage. We complement our experiments with tight binding simulations to show that the nature of the superconductor-graphene interface plays an important role in determining the CPR. We believe that the simplicity of our device architecture and measurement scheme should make it possible to use such devices for studies of the CPR in topologically non-trivial graphene Josephson junctions.

\emph{Acknowledgements}: We thanks A. Geresdi and D. van Woerkom for useful discussion. S.G. and L.M.K.V acknowledge support from the EC-FET Graphene flagship and the Dutch Science Foundation NWO/FOM. A.K. acknowledges funding from FLAG-ERA through project 'iSpinText'. P.R. acknowledges the support of the OTKA through the grant K108676, the support of the postdoctoral research program 2015 and the support of the J\'anos Bolyai Research Scholarship of the Hungarian Academy of Sciences. K.W. and T.T. acknowledge support from the Elemental Strategy Initiative conducted by the MEXT, Japan and JSPS KAKENHI Grant Numbers JP26248061,JP15K21722 and JP25106006.

\clearpage
\setcounter{figure}{0}
\renewcommand{\thefigure}{S\arabic{figure}}
\onecolumngrid
\section{\large{Supplementary Information}}
\vspace{10mm}

\section{1. Device Fabrication}

Graphene flakes are exfoliated onto silicon chips with 90~nm SiO$_2$. Next, h-BN is exfoliated separately on a glass slide covered by a 1-cm$^{2}$ piece of PDMS coated with an MMA/MAA copolymer layer. This glass slide is baked for 20 minutes on a hot plate at $120 ^{\circ}\mathrm{C}$, prior to h-BN exfoliation. The glass slide is mounted on a micromanipulatior in a home-built set-up (similar to Ref~\cite{Steele_s}) equipped with a heating stage. Next, a h-BN flake on the slide is aligned with the target graphene and the substrate is heated to $90^{\circ}$C. The flakes are brought into contact, after which the glass slide is released smoothly such that the graphene flake is picked up by the h-BN flake on the glass slide. Finally, the graphene/h-BN stack is transferred onto another h-BN flake (exfoliated onto a silicon chip with 285~nm SiO$_2$), at a temperature of $80^{\circ}$C. 

\begin{figure}[!h]
	\renewcommand*{\thefigure}{S\arabic{figure}}
	\includegraphics[width=.8\linewidth]{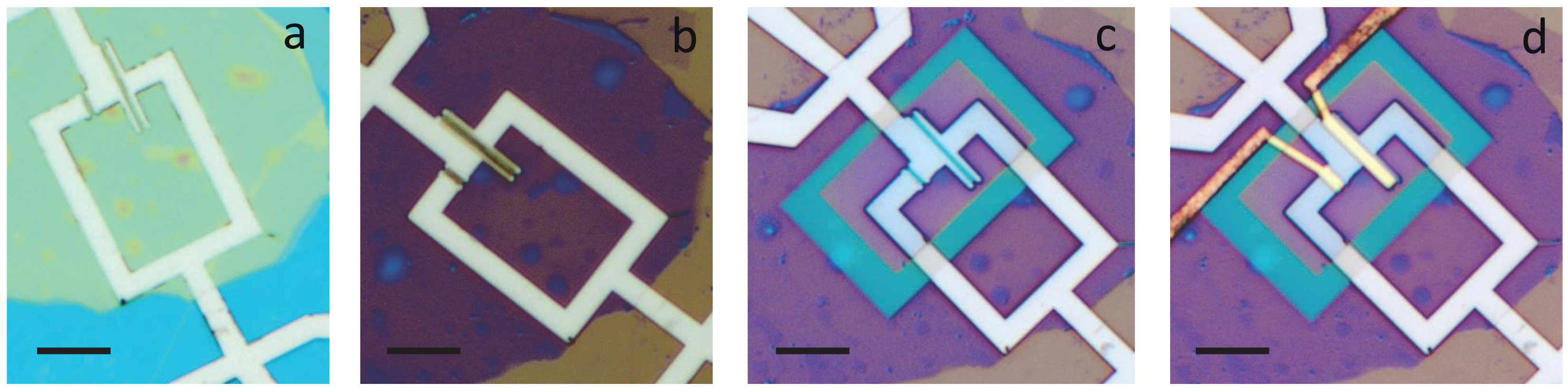}
	\caption{Optical images of device (Dev2 in main text) after (a) MoRe deposition, (b) shaping of the graphene, (c) dielectric (HSQ) deposition, and (d) deposition of top gates. The scale bar for all images is 5~$\mu$m.} 
	\label{fig:fabrication}
\end{figure}

The processing flow is outlined in Figure~\ref{fig:fabrication}. First MoRe contacts are made to the stack via an etch fill technique~\cite{Calado_s} using standard e-beam lithography. The sample is plasma-etched for 1~min in a flow of 40/4~sccm CHF$_{3}$/O$_{2}$ with 60~W power, and $80\mu$bar pressure. After etching, we immediately sputter $\sim$70~nm MoRe using a DC plasma with a power of 100~W in an Argon atmosphere. Next, the MoRe lift-off is completed in hot ($54 ^{\circ}\mathrm{C}$) acetone for about 3-4 hours. The two JJs are shaped using another e-beam lithography in which the intended graphene geometry is defined via a PMMA/hydrogen-silsesquioxane (HSQ) mask, followed by CHF$_{3}$/O$_{2}$ etching. In order to isolate the contacts from the top gate, we use~$\sim$170~nm of HSQ as a dielectric. Finally, top gates are fabricated by e-beam evaporation of 5nm Cr + 120 nm Au, and subsequent lift off in hot acetone.
 
\section{2. Ballistic transport in Dev2}

\begin{figure}[!h]
	\renewcommand*{\thefigure}{S\arabic{figure}}
	\centering
	\includegraphics[width=.8\linewidth]{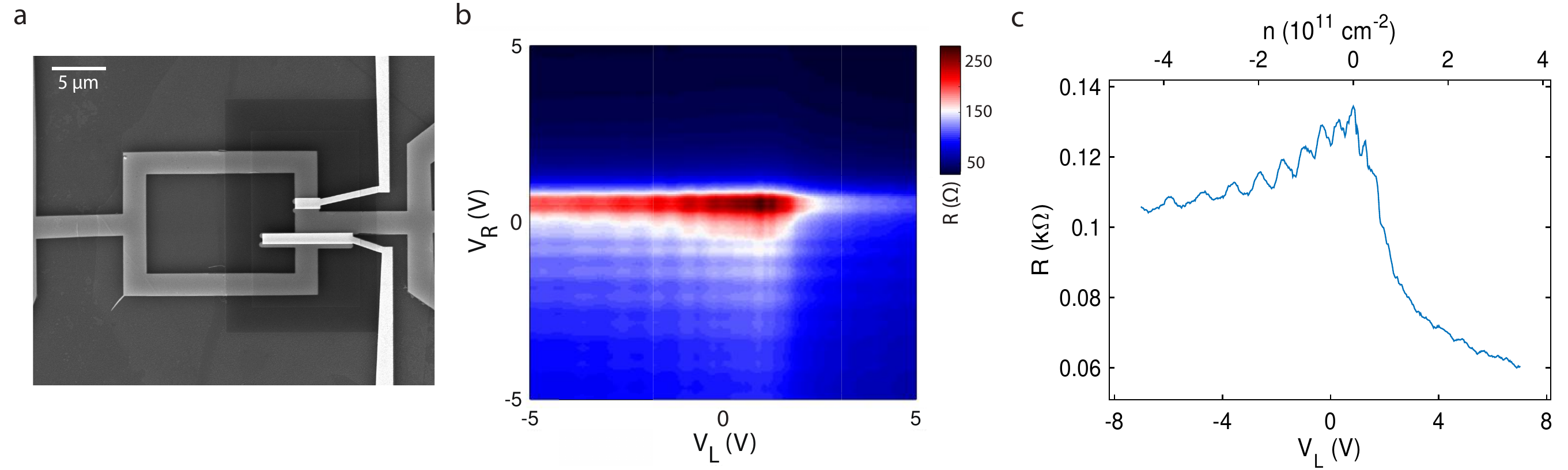}
	\caption{(a) Scanning electron micrograph of Dev2 from main text. The left junction ($L_{JJ}$) is 0.4~$\mu$m long (L) and 8~$\mu$m wide (W), while the right junction ($R_{JJ}$) is 0.4~$\mu$m long and 2~$\mu$m wide. (b) Resistance map as a function of $V_{L}$ and $V_{R}$ at 4.2~K, demonstrating independent control of carrier type and density in left and right JJ, respectively. (c) Resistance vs $V_{L}$ (while keeping $V_{R}$ fixed at CNP of R-JJ) showing Fabry-P\'erot oscillations in resistance.}
	\label{fig:Dev2} 
\end{figure}

\newpage
\section{3. Magnetic field to phase conversion}

\begin{figure}[!b]
	\renewcommand*{\thefigure}{S\arabic{figure}}
	\centering
	\includegraphics[width=.6\linewidth]{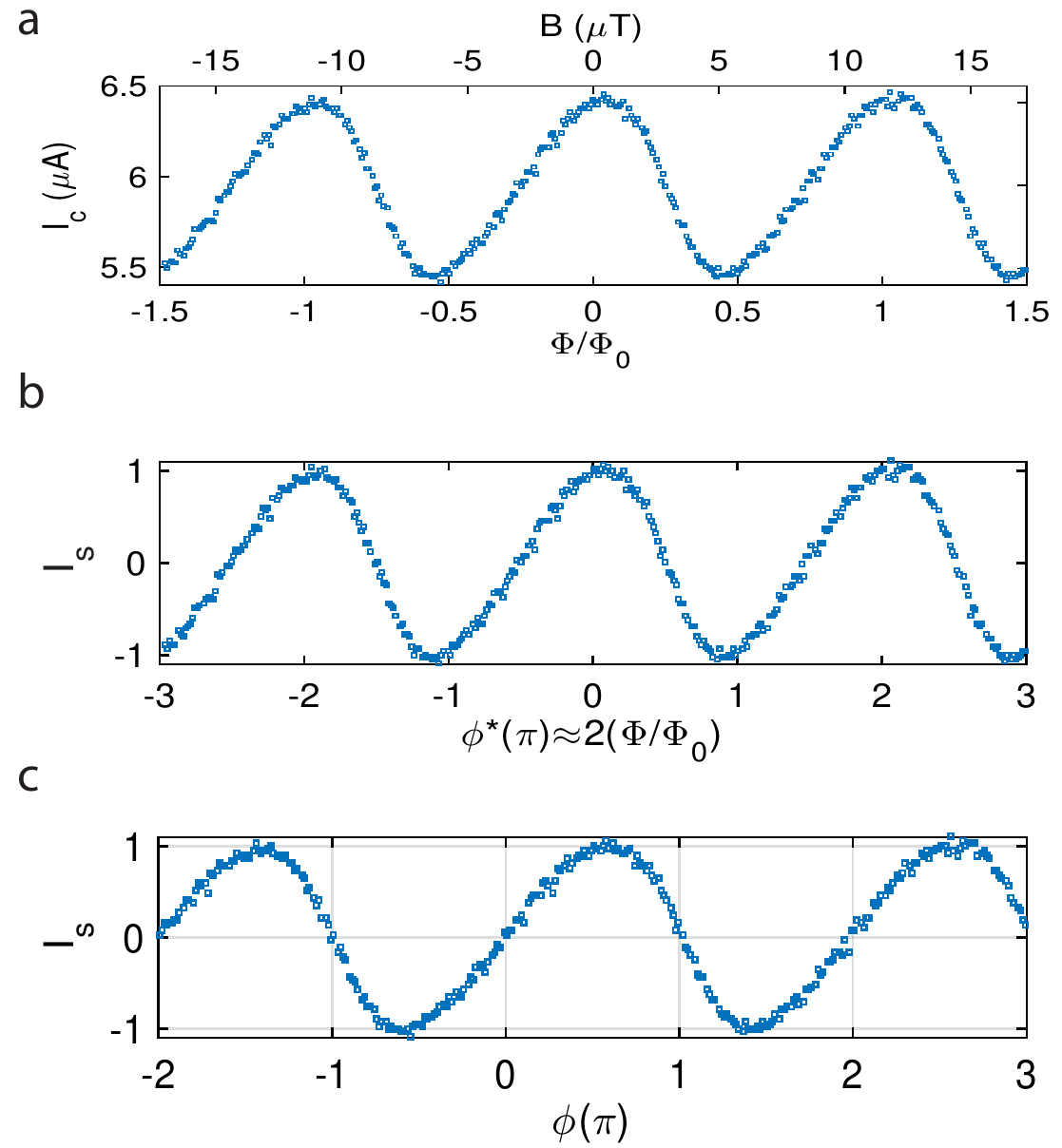}
	\caption{(a) The variation of $I_{c}$ as a function of magnetic field $B$ for $V_{L}=-4$~V and $V_{R}= +10$~V. (b) The curve in (a) replotted after converting flux $\Phi$ to phase $\phi^*$, and rescaling $I_c$ to $I_s = (I_c-I_{cR})/I_{cL}$. (c) Curve in (b) offset along the $\phi^*$-axis to ensure that $I_s (0) = 0$, thus giving the true phase $\phi$ axis.} 
	\label{fig:conversion}
\end{figure}

In the main text we pointed out that one must take care in converting the flux-periodic oscillations of the critical current of the SQUID $I_c(\Phi)$ to the CPR of L-JJ $I_s(\phi)$. Figure~\ref{fig:conversion} shows how this is done. We start with the upper plot in Figure~2a of the main text, which is shown here again for convenience (Figure~\ref{fig:conversion}a). We then subtract a constant background ($I_{cR}$) about which the curve oscillates and normalize it with respect to the oscillation amplitude ($I_{cL}$). Also, the flux is converted to phase by $\phi^* \rightarrow 2\Phi/\Phi_0$. This is not the true phase $\phi$ for two important reasons. Firstly, the zero of the magnetic field is not known precisely. Secondly, the flux to phase conversion is only possible up to a constant offset, which is determined by the CPR of R-JJ (which is a-priori unknown). In order obtain the CPR we then offset the curve in Figure~\ref{fig:conversion}b along the $\phi^*$ axis such that the supercurrent at zero phase difference is zero, which finally gives us the CPR. We note that this procedure is only valid for systems where $I_s(\phi) = -I_s(-\phi)$ and $I_s (0) = 0$, both of which are reasonable assumptions for our graphene JJs.       

\newpage

\section{4. Eliminating inductance effects}
In an asymmetric SQUID inductance effects can give rise to skewed $I_c (\Phi)$ curves. It is therefore important to establish that such effects do not dominate the response of the SQUIDs described in this study. To do so, we first provide some qualitative arguments which make it evident that the skewness arises only from a non-sinusoidal CPR. Furthermore, we extract the loop inductance of our SQUID, use it as an input for the RCSJ model and confirm that (within our experimental resolution) the inductance does not play an important role in determining the shape of the $I_c (\Phi)$ curves, and hence does not affect our ability to measure the CPR.

\subsection{Large asymmetry}

\begin{figure}[!b]
	\renewcommand*{\thefigure}{S\arabic{figure}}
	\includegraphics[width=.65\linewidth]{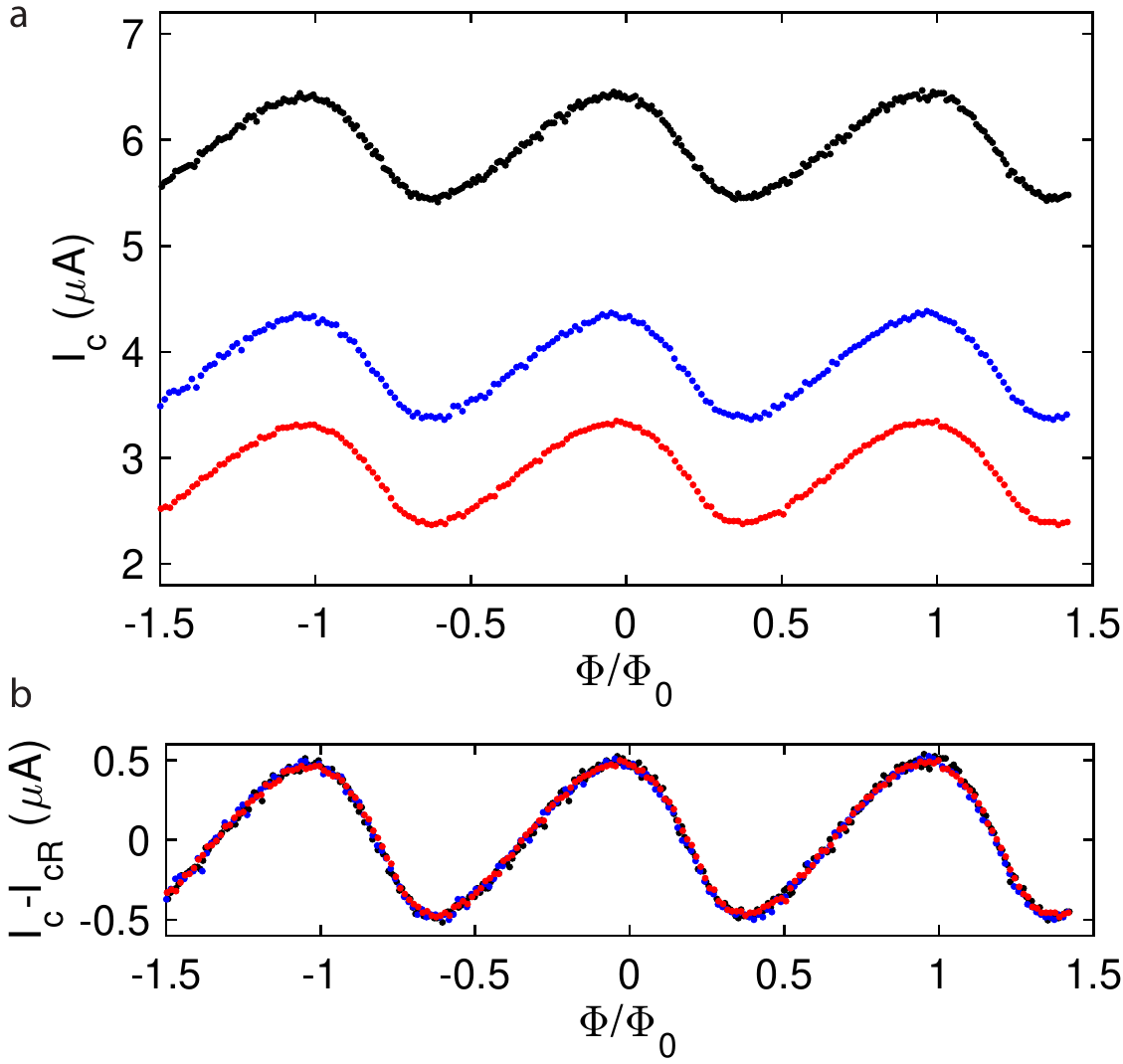}
	\caption{(a) $I_{c}(\Phi)$ plots with fixed $I_{cL}$, but varying $I_{cR}$, as shown earlier in Figure~2a of the main text. (b) The curves in (a), now plotted as $(I_{c}-I_{cR})$ vs. $\Phi$. Collapse of the curves shows that the skewness does not depend on $I_c$, and hence represents the CPR of L-JJ.} 
	\label{fig:Inductance1}
\end{figure}

We have shown that for large asymmetry (i.e., $I_{cR}>>I_{cL}$), we probe the CPR of L-JJ. We define the asymmetry parameter $a_i = (I_{cR}-I_{cL})/(I_{cR}+I_{cL})$. Figure \ref{fig:Inductance1}a shows three traces at $T=40$~mK, where $I_{cL}\approx 0.5$~$\mu$A is kept fixed and $I_{cR}$ is varied from 6~$\mu$A (black trace, $a_i\approx 0.83$) to 2.8~$\mu$A (red trace, $a_i\approx 0.78$). Figure \ref{fig:Inductance1}b shows that all three curves collapse despite the fact that the maximum critical current ($I_{max} = I_{cR}+I_{cL}$) changes by a factor of two. If the skewness was dominated by inductance effects, we would have not expected this collapse, since the screening parameter $\beta_L = I_{max}L/\Phi_0$ increase by a factor of two (going from the red trace to the blue trace). In other words, the combined effect of large asymmetry and inductance should have resulted in a larger skewing of the black trace (maximum $\beta_L$ and $a_i$) as compared to the red one.

\subsection{Intermediate asymmetry}

\begin{figure}[!t]
	\renewcommand*{\thefigure}{S\arabic{figure}}
	\includegraphics[width=.6\linewidth]{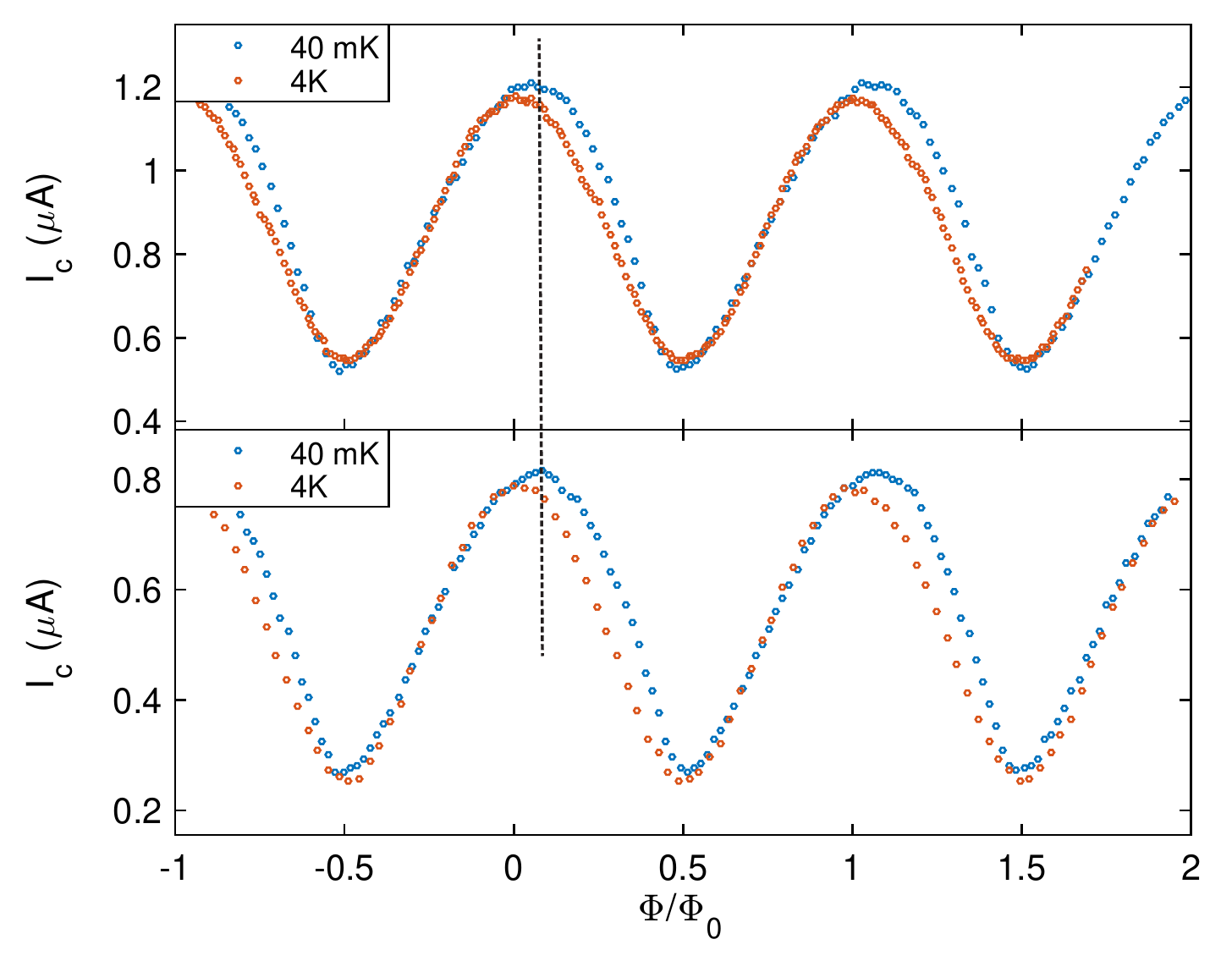}
	\caption{$I_{c}(\Phi)$ plots at $I_{max} \approx 1.2$~$\mu$A (upper) and $\approx 0.8$~$\mu$A (lower) for low asymmetries of $a_i = 0.45$, $a_i = 0.33$ respectively. The curves at 40~mK are skewed (indicated by position of dashed line), while those at 4.2~K are not.}
	
	\label{fig:Inductance2}
\end{figure}

We have shown in the main text (Figure~5) that the skewness of the CPR decreases with increasing temperature, resulting in a sinusoidal CPR at 4.2~K. One might argue that this is consistent with inductance effects, whereby an increase in temperature reduces the critical currents and hence $\beta_L$. To eliminate this possibility, we compare $I_c (\Phi)$ at 40~mK and 4.2~K. Figure \ref{fig:Inductance2}a,b show two such data sets. In each case the gate voltages were tuned such that both $I_{max}$ and $a_i$ were roughly the same for both temperatures. We see that at 40~mK the curves are noticeably skewed as compared to 4.2~K. The asymmetry here is not sufficient to directly extract the CPR, but it clearly demonstrates that the non-sinusoidal CPR also manifests itself in skewed $I_c (\Phi)$ curves at intermediate asymmetry. We note that this argument is made stronger by the fact that the inductance at 4.2~K should in fact be larger than that at 40~mK, since the inductance of the MoRe loop is dominated by kinetic inductance, which increases at higher temperatures. In other words, one would expect inductance related effects to be enhanced at higher temperatures, rather than become suppressed.  

\subsection{Estimating the loop inductance}

\begin{figure}[!t]
	\renewcommand*{\thefigure}{S\arabic{figure}}
	
	\includegraphics[width=1\linewidth]{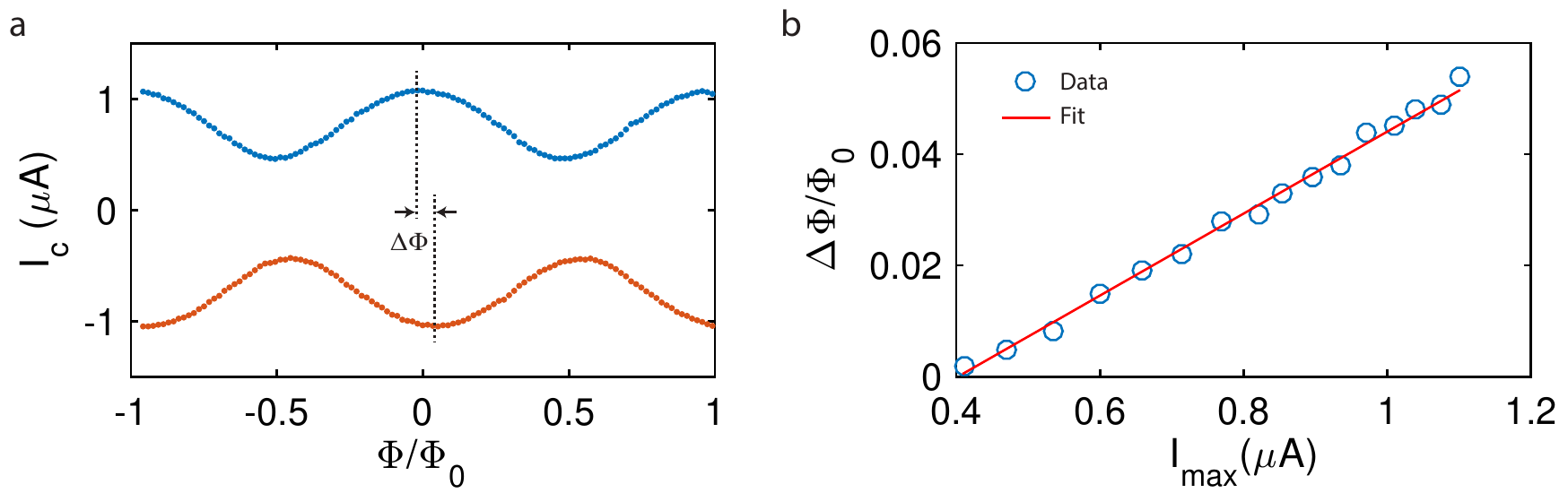}
	\caption{(a) $I_{c}(\Phi)$ curves for an asymmetric SQUID. (b) Variation of $\Delta\Phi$ with $I_{max}$. Here $I_{cL}$ is kept fixed, while $I_{cR}$ is varied. Blue circles are the experimentally obtained values of $\Delta\Phi$ and the red line is a linear fit to the data.}
	\label{fig:Inductance3}
	
\end{figure}

Figure \ref{fig:Inductance3}a shows $I_c(\Phi)$ measurements of an asymmetric SQUID at 4.2~K, where we have established that the CPR is sinusoidal. The position of maximum $I_c$ for positive and negative current bias are offset along the flux axis due to self-flux effects~\cite{Goswami_s}. This shift is given by: $\Delta\Phi = 2L(I_{cR}-I_{cL})$, where $I_{cR}$ and $I_{cL}$ are the critical current of right and left junction respectively. Figure \ref{fig:Inductance3}b shows the variation of $\Delta\Phi$ with $I_{max}$. These values are obtained by keeping $I_{cL}\approx 0.2$~$\mu$A fixed and varying $I_{cR}$ from 0.2~$\mu$A (symmetric configuration) to 0.9~$\mu$A (highly asymmetric). Since $\partial\Delta\Phi / \partial I_{cR} = L$, a linear fit (red line) allow us to extract $L\approx 152$~pH. Since MoRe is a highly disordered superconductor, its inductance is dominated by the kinetic inductance and the low temperature inductance $L (0) = L (T) [1-(T/T_c)^2]$, giving $L\approx 110$~pH at $T=40$~mK. We use this inductance to compare our experiments with the RCSJ simulations described below. 

\section{5. RCSJ Model}

\begin{figure}[!h]
	\renewcommand*{\thefigure}{S\arabic{figure}}
	\centering
	\includegraphics[width=.7\linewidth]{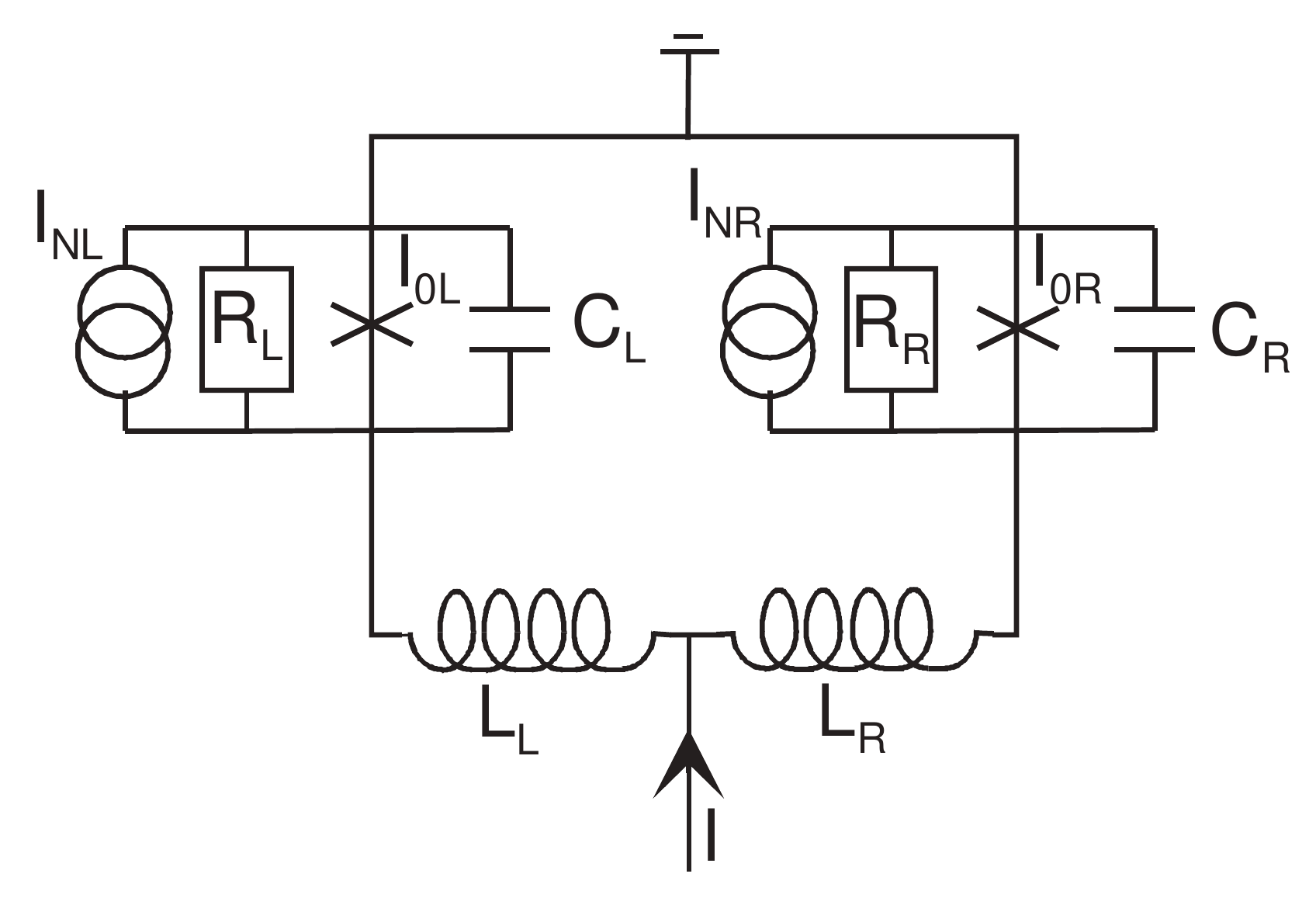}
	\caption{Circuit diagram of the asymmetric SQUID.}
	\label{fig:RCSJFigure} 
\end{figure}

To model the asymmetric SQUID we consider the circuit shown in Figure \ref{fig:RCSJFigure}. The Josephson junctions are described by the resistively and capacitively shunted junction (RCSJ) model\cite{McCumber_s, Stewart_s} by Josephson currents with phases $\delta_{L}$ and $\delta_{R}$ and amplitudes $I_{cL}= I_{c} (1-a_{i})$  and $I_{cR}= I_{c} (1+a_{i})$, resistors $R_{L}$ and $R_{R}$ , and capacitors $C_L$ and $C_R$. The Josephson currents are given by $I_{L}= I_{c} (1-a_{i}) \cdot f_L(\delta_{L})$  and  $I_{R}= I_{c} (1+a_{i}) \cdot f_L(\delta_{L})$, where $f_{i}(\delta_{i})$ are the normalized current-phase relations of the left and right JJ, respectively. 
The Nyquist noise arising from the two resistors is described by two independent current noise sources $I_{NL}$ and $I_{NR}$ having white spectral power densities 4$k_{B}T/R_{L}$ and 4$k_{B}T/R_{R}$, respectively. The two arms of the SQUID loop have inductances $L_{L}$  and $L_{R}$ . The total inductance $L$ is the sum of the geometric (\textit{L$_{g}$}) and the kinetic (\textit{L$_{k}$}) inductance. The loop is biased with a current $I$, and a flux $\Phi$ is applied to the loop. 

\newpage

In the following we are interested only in static solutions and normalize currents by $I_{c}$. The currents  $i_{L}$ and $i_{R}$ through the left (right) arm of the SQUID are then given by:

\begin{equation}
i_{R}=(1+a_{i}) \cdot f_R(\delta_{R}).
\label{eq:RCSJ1}
\end{equation}

\begin{equation}
i_{L}=(1-a_{i}) \cdot f_L(\delta_{L}).
\label{eq:RCSJ2}
\end{equation}

Assuming for simplicity that $L_{L} = L_{R}$ (a reasonable assumption based on our device geometry) the normalized circulating current $\textit{j}$ is given by:

\begin{equation}
j= \frac{i_{R}-i_{L}}{2}= \frac{1}{\beta_{L}}\Big(\frac{\delta_{L}-\delta_{R}}{\pi}-2 \Phi/\Phi_0 \Big).
\label{eq:RCSJ3}
\end{equation}

and the maximum current across the SQUID is $i_{cR} + i_{cL}$. From Equation \ref{eq:RCSJ3} we obtain

\begin{equation}
\delta_{L}=2 \pi \Phi/\Phi_0 + \delta_{R} + \pi \beta_{L} \frac{i_{R}-i_{L}}{2}.
\label{eq:RCSJ4}
\end{equation}

Let us consider the case $a_i>>0$, i.e., the right junction has a much larger critical current than the left one. As we will see, in this case the modulation of the SQUID critical current reflects the CPR of the left junction, provided that \textit{$\beta_{L} <<1$}.  

\begin{equation}
i=i_{R}+i_{L}=(1+a_{I}) \cdot f_R(\delta_{R})+(1-a_{i}) \cdot f_L(\delta_{L}).
\label{eq:RCSJ5}
\end{equation}

From  Equation \ref{eq:RCSJ4}, for \textit{$\beta_{L}<<1$}, we obtain $\delta_{L} \approx 2 \pi \Phi/\Phi_0 + \delta_{R}$. Thus

\begin{equation}
i=i_{R}+i_{L}=(1+a_{I}) \cdot f_R(\delta_{R})+(1-a_{i}) \cdot f_L( 2 \pi \Phi/\Phi_0+\delta_{R}).\label{eq:RCSJ6}
\end{equation}

Let us assume that $i > 0$. Then the task is to maximize $i$ with respect to $\delta_{R}$, to obtain \textit{i$_{c,SQUID}$} vs $\Phi/\Phi_0$.
If the critical current of the right JJ is much bigger than the critical current of the left JJ, the value of $\delta_{R}$ will be close to the value $\delta_{R}^{0}$  where the CPR of the right JJ has its maximum. We thus Taylor expand: 

\begin{equation}
f_R(\delta_{R}) \approx f_R(\delta_{R}^{0}) + \frac{1}{2} \frac{d^{2}f_R}{d\delta_{R}^{2}} \bigg|_{\delta_{R}^{0}} (\delta_{R}-\delta_{R}^{0})^{2} + ....
\label{eq:RCSJ7}
\end{equation}

Note that in Equation \ref{eq:RCSJ7} the first derivative of $f_R$ is zero, because we look for the maximum of this CPR. 
If the second derivative $(< 0)$ is reasonably peaked,  $\delta_{R}$ will be very close to $\delta_{R}^{0}$  and we obtain:

\begin{equation}
i_{c,SQUID} \approx (1+a_{i}) \cdot f_R(\delta_{R}^{0}) + (1-a_{i}) \cdot f_L(2 \pi \Phi/\Phi_0 + \delta_{R}^{0}) = const + (1-a_{i})f_L(2 \pi \Phi/\Phi_0+ \delta_{R}^{0}).
\label{eq:RCSJ8}
\end{equation}

This means that \textit{i$_{c,SQUID}$} vs. $\Phi/\Phi_0$ probes the CPR of the left JJ up to  a phase shift  $\delta_{R}^{0}$ .  $f_L$ can be evaluated further if one assumes that $f_L = 0$ at $\delta_{L}= 0$  and that min($f_{L}$) = - max($f_L$).

\begin{figure}[!t]
	\renewcommand*{\thefigure}{S\arabic{figure}}

	\includegraphics[width=0.6\linewidth]{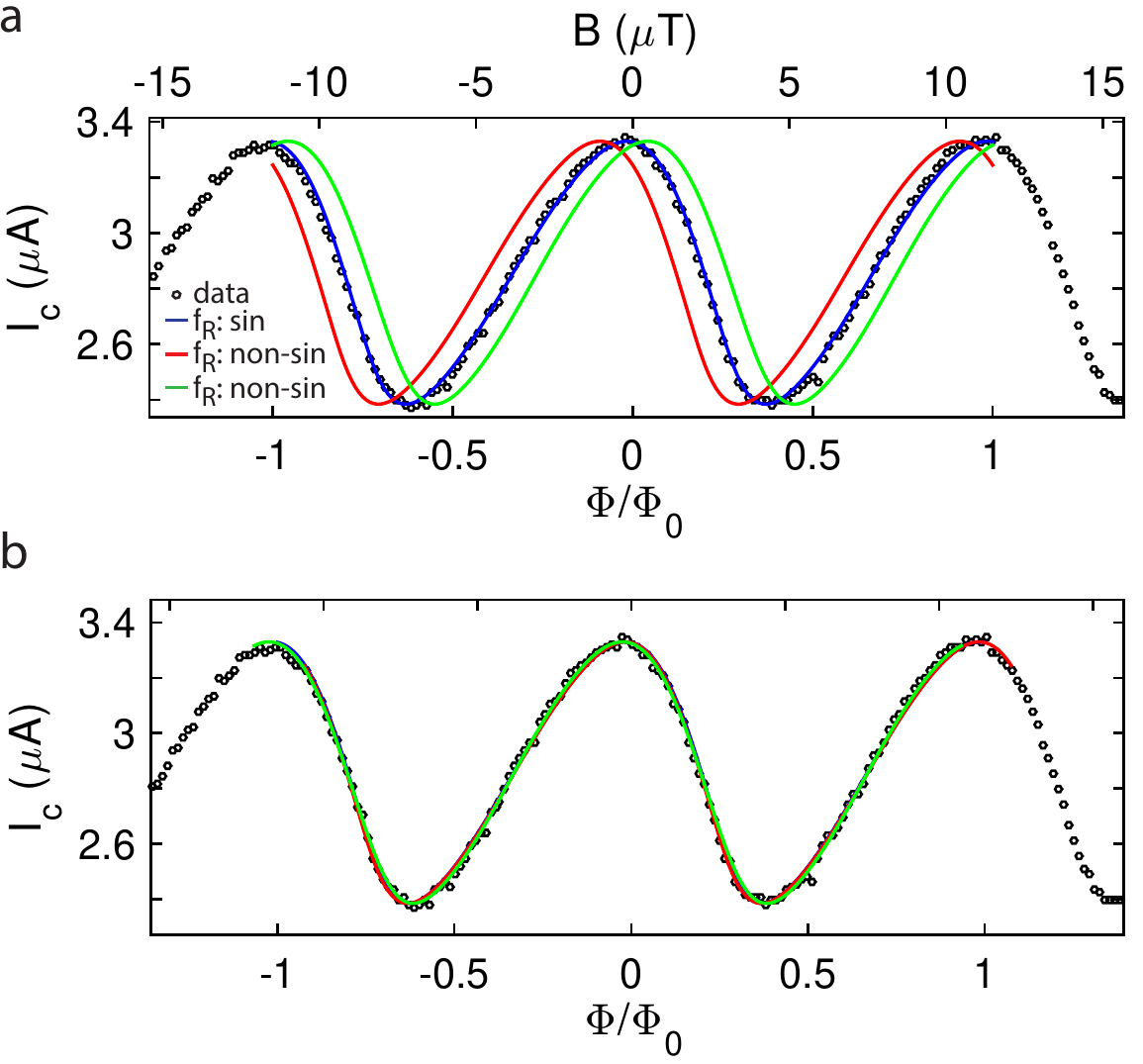}
	\caption{(a) Experimental $I_c (\Phi)$ (black) along with RCSJ simulations for a SQUID with $a_{i}=0.83$. The blue (red) curve corresponds to a sinusoidal (non-sinusoidal) CPR $f_R$ for the reference junction R-JJ, with the experimentally determined $\beta_L = 0.34$. The green curve shows the result for $\beta_L = 0.01$. The data has been offset along the flux axis to match the blue curve. (b) Same as (a), but with the red and green curves shifted along the flux axis.}
	\label{fig:Inductance4}
	
\end{figure}

In Figure~2a of the main text we have compared our experiments with a full RCSJ simulation, as described above. These simulations involve no free parameters since we use the experimentally determined inductance, asymmetry ($a_i$), and CPR of L-JJ $f_L (\delta_L)$  as input parameters. For simplicity, the numerical plots were generated assuming a sinusoidal CPR for the reference junction R-JJ, shown as the blue curve in the Figure~\ref{fig:Inductance4}a. The red curve shows how $I_c(\Phi)$ changes when R-JJ is assumed to have a non-sinusoidal CPR (with a functional form similar to that extracted for L-JJ). The only effect this has is to offset the simulated curves along the flux axis. This is a consequence of the fact that  $\delta_{R}^{0}$ (as described above) is different for the two cases. However, we see in Figure~\ref{fig:Inductance4}b that these two cases perfectly overlap with an appropriate offset along the flux axis. This confirms the fact that an incomplete knowledge of the CPR of R-JJ is (in practice) equivalent to an unknown offset in magnetic field, and therefore does not affect our ability to accurately determine the functional form of the CPR of L-JJ. The green curve in Figure~\ref{fig:Inductance4}a is a simulation with $\beta_L = 0.01$ (i.e., in the limit where the loop inductance is negligible). Looking carefully at Figure~\ref{fig:Inductance4}b shows that this $I_c(\Phi)$ has a slightly different shape as compared to the blue/red curves. However, this difference is well within the error bars for our estimation of the skewness, and we can conclude that the functional form of the $I_c(\Phi)$ curves is not dominated by the inductance effects, but by the non-sinusoidal CPR of L-JJ. This is in agreement with the conclusions drawn from more qualitative arguments presented in the previous section.  

\newpage
\section{6. Tight Binding-Bogoliubov-de Gennes Calculations}

\subsection{Details of the theoretical model}

In this Section we provide further details of the theoretical model that we used in our numerical 
calculations. As it will be clear from the following discussion, we found that in order to obtain
a  good qualitative agreement with the measurements, a realistic and detailed model of the 
Josephson junction, 
especially the interface between the superconductor and the normal regions, is  needed. 

We assume  that the  graphene flake which serves as a weak link is perfectly ballistic and 
scattering processes only occur at the interfaces between regions of different doping in the normal
part of the junction or between the superconductor and the normal region. 
The normal (N) and superconducting (S) regions are of the same width in our calculations. 
This  allows us to use periodic boundary conditions perpendicular to the transport direction. 
The transverse momentum $q_n$ is 
a good quantum number and the DC Josephson current can be calculated as a sum over all $q_n$:
\begin{equation}
I_J(\Delta\phi) = \sum\limits_{n} I_J(q_n,\Delta\phi) \;,
\end{equation}
where  $I_J(q_n,\Delta\phi)$ is the momentum resolved Josephson current calculated for a specific transverse momentum $q_n$ 
via the contour integral method  developed recently in Reference \cite{method_s}. 
For wide junctions and high dopings, when there are many transverse momenta, 
the exact form of the boundary conditions is not important and therefore we 
used the infinite mass boundary condition to obtain $q_n$:
$
q_n = \left(n+\frac{1}{2}\right)\frac{\pi}{W}\;,
$
where $n=0,1,2,\dots$ and $W$ is the width of the junction.

\begin{figure}[!t]
	\centering
	\renewcommand*{\thefigure}{S\arabic{figure}}
	\includegraphics[scale=0.75]{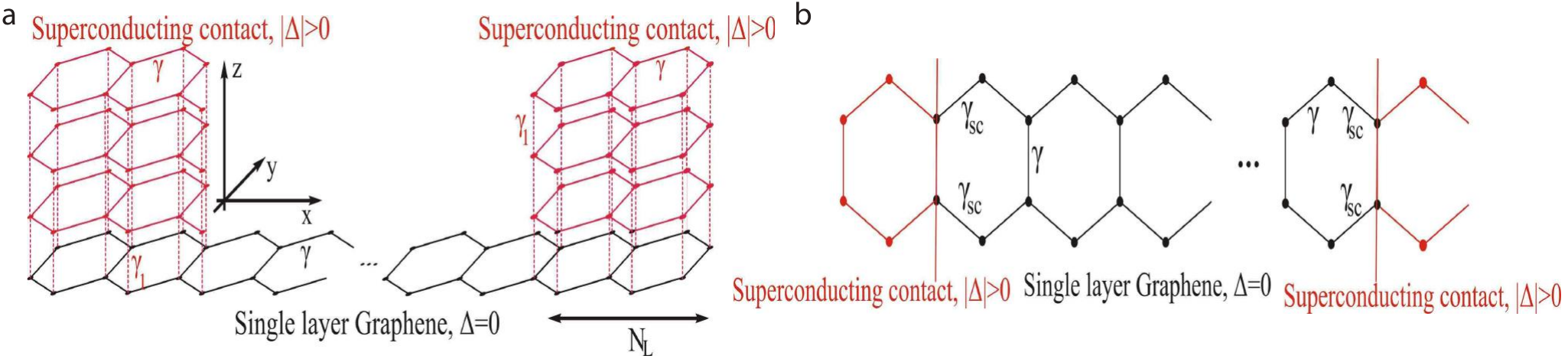}
	\caption{ (a) The geometry of the top contacted superconductor-graphene-superconductor junction. 
		$N_L$ is the number of unit cells under the superconducting contacts in the $x$ direction, (b) The side-contacted geometry. The interface resistance between the S and regions is modelled 
		by a hopping $\gamma_{sc}<\gamma$. In both geometries the lattice is translational invariant in the $y$ direction} 
	\label{fig:contact-geometries}
\end{figure}

The description of both the N region and the S terminals is based on the nearest-neighbour tight-binding 
model of graphene\cite{graphene-TB_s} 
\begin{equation}
\hat{H} = \sum_{i} U_i^{} c_{i}^{\dagger} c_i - \sum_{\langle ij\rangle}\gamma_{}^{}c_{i}^{\dagger} c_j+ h.c.
\end{equation}
where $U_i$ 
is the on-site energy on the atomic site $i$, $\gamma_{}=2.97\,$eV is the
hopping amplitude between the nearest-neighbor atomic sites 
$\langle ij \rangle $ in the graphene lattice, and $c_i^{\dagger}$ ($c_i$ ) is a creation (annihilation) operator 
for electrons at site $i$. 
We  considered two junction geometries. Most of our results  were obtained using the 
top-contact geometry, which is shown in Figure 4(a) of the main text and for convenience 
repeated here in Figure \ref{fig:contact-geometries}(a).  The S terminals are described by vertically stacked 
graphene layers (AA stacking) where the inter-layer hopping is given by $\gamma_1=0.6\,$eV.  
The same inter-layer hopping $\gamma_1$ was also used between the S terminals and the N region. 
The S leads  are coupled to the normal graphene sheet over $N_{L}$ unit cells. The 
result do not depend strongly on the exact value of $N_{L}$, therefore we used $N_{L}=10$ in our calculations. 

To mimic  metallic leads with many open channels, the S terminals are highly  $n$-doped. 
This is described by an on-site potential  $U_n$ and  we used  $U_n=350\,$meV in our calculations. 
For high $n$-doping of the N region we calculated  an average transparency of
$Tr=0.82$ for the junction, see  the Supplementary of Reference \cite{Falko_s} for the precise 
definition of  $Tr$. We find that the calculated $Tr$ does not depend very sensitively on the 
precise value $U_n$ and  $\gamma_1$  because most of the backscattering taking place at the interface 
of the S leads and the N region is due to a  ``geometric'' effect: the electron trajectories have  to turn 
at right angle to arrive from the lead into the N region. 
Moreover, we find that for  $Tr=0.82$ the calculated dependence of the normal 
state resistance $R_n$ on the doping of the N region  agrees qualitatively well with the measurements where 
the right JJ was kept at the charge neutrality point
[c.f. Figure 1(c) in the main text and Figure \ref{fig:side-contact}(a) below].
(We did not try to achieve quantitative agreement for $R_n$ because in the experiments the resistance of the two
junctions are always measured in parallel, whereas we used single junctions in the calculations.) 

As shown in Figure \ref{fig:contact-geometries}(b) and discussed further later on,
we have also made calculations using the  side-contact geometry. For both geometries we used open boundary conditions 
for the leads in the transport direction 
(which is the $z$ direction in top-contacted geometry and the $x$ direction in the side-contacted one, 
see Figure \ref{fig:contact-geometries}).

In contrast to the S leads, which are always n-doped in our calculations, 
the normal region of the JJ can be either $n$ or $p$ doped depending on the gate voltage. 
This is modeled  by a  doping potential $U_p$. 
%
Experimentally, it was shown that the {superconducting terminals}  $n$-dope the normal region of the JJ \cite{Calado_s,Falko_s}.  
This $n$-doped region extends to a distance $x_1$ ($L_0-x_2$) from the left (right) terminal, 
where $L_0$ is the distance between the two S leads.
The  potential profile in the junction can be therefore either $npn$ or $nn'n$.
The exact value of the $x_1$ and $x_2$, and hence the cavity length $L^{*}=x_2-x_1$, however,  
depends on the gating of the JJ.
In the $npn$ regime, where clear FP oscillations can be measured in the normal state
resistance $R_n$ in our devices, we  extracted  the experimental cavity length using the relation 
$L_{exp}^{*}\approx 2\sqrt{\pi n}/\delta n$, 
where $\delta n$ is the density difference between consecutive peaks in $R_n$~\cite{rickhaus_s}. 

\begin{figure}[!t]
	\centering
	\includegraphics[width=9cm]{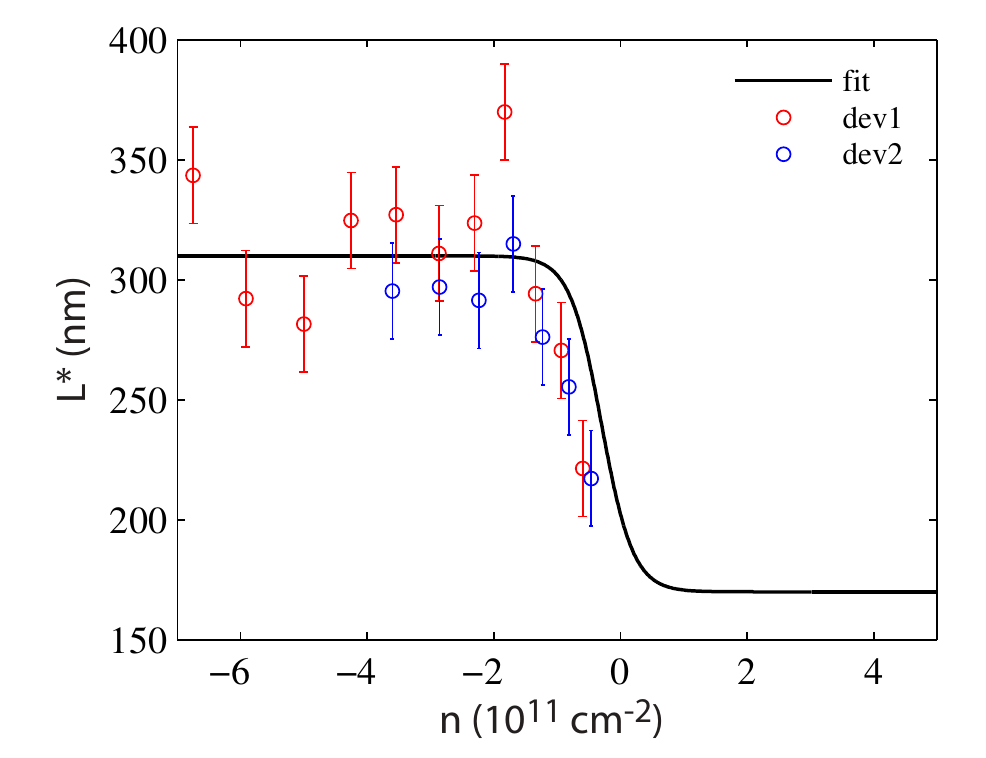}
	\caption{The experimental cavity length $L_{exp}^{*}$ vs doping (symbols) and the fitting function used to obtain the 
		$L^{*}$ in our calculations (solid line).} 
	\label{fig:L*}
\end{figure} 
The results of this analysis are summarized in Figure~\ref{fig:L*}. We find that $L_{exp}^{*}\approx 310$\,nm 
is roughly constant for $n<-1.8\times10^{11}{\rm cm}^{-2}$, but it decreases for densities approaching  the CNP. 
In order to extract the theoretical cavity length $L^{*}$ for $n>-1.8\times10^{11}{\rm cm}^{-2}$, 
we fitted the experimental results by the function
\begin{equation}
L^*(U_p) = \frac{L_{exp}^{*}-L_{nn'n}}{1+\exp{[\beta (n-n_0)]}} +L_{nn'n}.
\label{eq:Lstar}
\end{equation}
Here $L_{nn'n}$ is the cavity length for strongly $n$ doped junctions which could not be determined 
from the $R_n$ measurements, therefore we used  $L_{nn'n}=170$\,nm. As mentioned above, a good qualitative 
agreement between the calulated and measured normal state resistance is achieved 
using this value of $L_{nn'n}$. We have also checked that  for  $L_{nn'n}\gtrsim 160$\,nm  the calculation 
results do not depend strongly on the exact value of $L_{nn'n}$. 
The two fitting parameters in Eq.~(\ref{eq:Lstar}) are  $\beta$ and $n_0$ 
and we found $\beta=4.0$ and $n_0=-0.3$, see Figure \ref{fig:L*}. 
Once $L^{*}$ is determined,  the parameters $x_1$ and $x_2$ are given by $x_1=\frac{L_0-L^*(U_p)}{2}$ and $x_2=L_0-x_1$. 
The total potential profile along the junction, which describes the smooth transition between the highly doped regions 
($x<x_1$ and $x>x_2$)  and the central part of the junction ($x_1\le x \le x_2$) is modeled by 
\begin{equation}
U(x) = U_n + \frac{U_p-U_n}{2} \left( \tanh\left(\frac{x-x_1}{l_{tr}}\right) - \tanh\left(\frac{x-x_2}{l_{tr}}\right) \right). 
\label{eq:potential}
\end{equation}
where the parameter $l_{tr}$ controls the smoothness of the transition. 
We used $l_{tr}=\frac{2}{5} x_1$ in our calculations corresponding to a relatively sharp 
transition. Larger values of $l_{tr}$ would effectively mean that the leads $n$ dope the N region of the 
junction and the doping there would therefore not be determined by $U_p$.

Finally, superconductivity in the S terminals is modelled by a on-site, complex
pair-potential $\Delta$ which goes to zero as a step-function at the S-N interface.
We made sure that that the Fermi-wavelengths $\lambda_N$ and $\lambda_S$ in the N and S  regions, respectively,  
satisfy $\lambda_S\ll \lambda_N$. This ensures that the exact spatial dependence of 
the superconducting pair potential at the N-S interface is not very important in the calculations\cite{beenakker-sgs_s}.


\subsection{Soft vs hard superconducting gap}

Following Reference \cite{soft_gap_s}, we also considered the effect of quasiparticle broadening in the 
superconducting terminals  by introducing a complex energy shift $E \rightarrow E + \rm{i} \eta$ 
in the self-energy calculations. 
Such a broadening, described by the parameter $\eta$,  can arise  due to scattering 
with phonons or other electrons or due to other effects leading to quasiparticle poisoning.

\begin{figure}[!t]
	\centering
	\renewcommand*{\thefigure}{S\arabic{figure}}
	\includegraphics[scale=0.8]{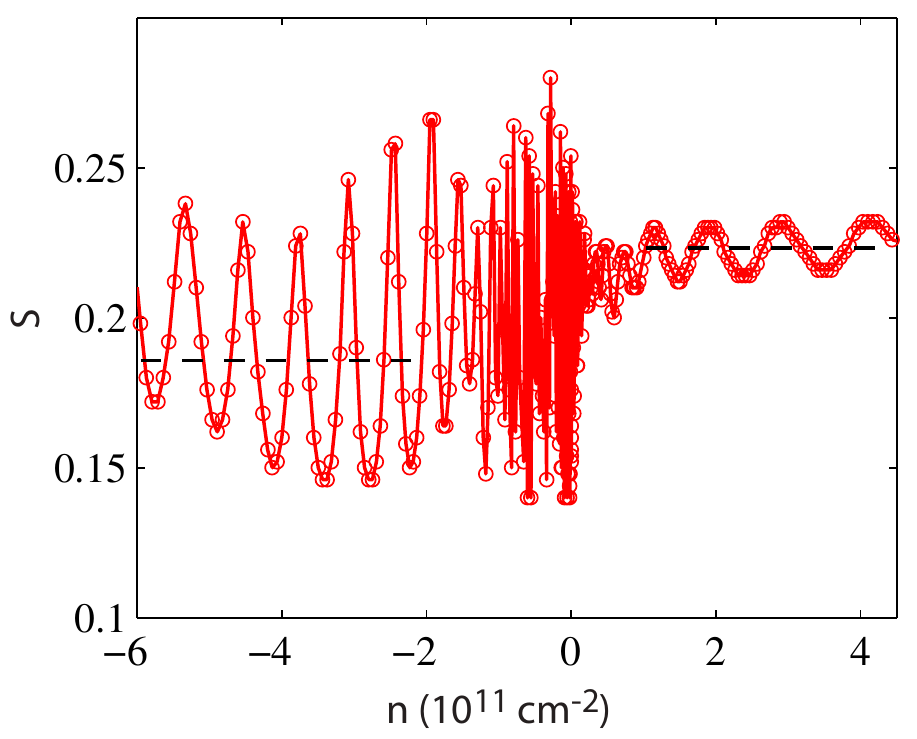}
	\caption{The calculated skewness vs doping for soft superconducting gap, i.e., $\eta=0.17\Delta$.} 
	\label{fig:skewness-eta}
\end{figure}

We find that a finite $\eta$ can considerably affect the value of the calculated 
critical current $I_c$. 
Since  $I_c$ is not the main focus of this work, we do not discuss the details here. 
Instead, we present results to illustrate the effect of $\eta$ on the skewness. 
We repeated the calculations using 
$\eta=0.17\Delta$ and the results are shown in Figure \ref{fig:skewness-eta}. 
Comparing Figure 4(d) in the main text and Figure \ref{fig:skewness-eta}, one can notice that 
the results are qualitatively very similar, but for $\eta=0$ the average skewness is larger 
for both  $npn$ and $nn'n$ doping than for $\eta\neq 0$. 
We note that  in the $nn'n$ regime the calculated average skewness $\bar{S}_{}=0.27$ for $\eta=0$ 
is closer to the measured one $S_{exp}\approx 0.28$ than the result $\bar{S}=0.22$ for $\eta=0.17\Delta$. 
The opposite is true in the $npn$ regime, where  the calculations with 
$\eta=0.17\Delta$ ($\eta=0$)  yielding $\bar{S}=0.19$  ($\bar{S}=0.22$) give better agreement with 
the measurements ($S_{exp}\approx 0.2$).  
We were not able to achieve an equally good agreement in both the $npn$ and $nn'n$ regimes 
using a single value of $\eta$. This may indicate that $\eta$  depends on the doping of the junction, 
but one would need a more microscopic understanding of the processes that contribute to $\eta$.  

We emphasize, however,  that $\eta$ is not the only parameter which can affect the value of the skewness.
Generally, the value of the skewness  depends on the interface between the S and  N regions. 
Calculations not shown here indicated that the presence/absence of a  
smooth transition between the highly doped leads and the normal graphene region 
(the parameter $l_{tr}$ in Eq.\ref{eq:potential}) and the value of the hopping amplitude  
$\gamma_{sc}$ in Figure \ref{fig:contact-geometries}(b) between the S and N regions can also 
affect the results. However, we fixed the value of the parameters describing the junction 
such that we obtain a qualitatively good agreement for $R_n$ (as discussed previously) 
and did not  changed these parameters in the skewness calculations.


\subsection{Calculations using the  side contact geometry}
\label{subsec:side-contact}

\begin{figure}[!b]
	\centering
	\renewcommand*{\thefigure}{S\arabic{figure}}
	\includegraphics[scale=0.7]{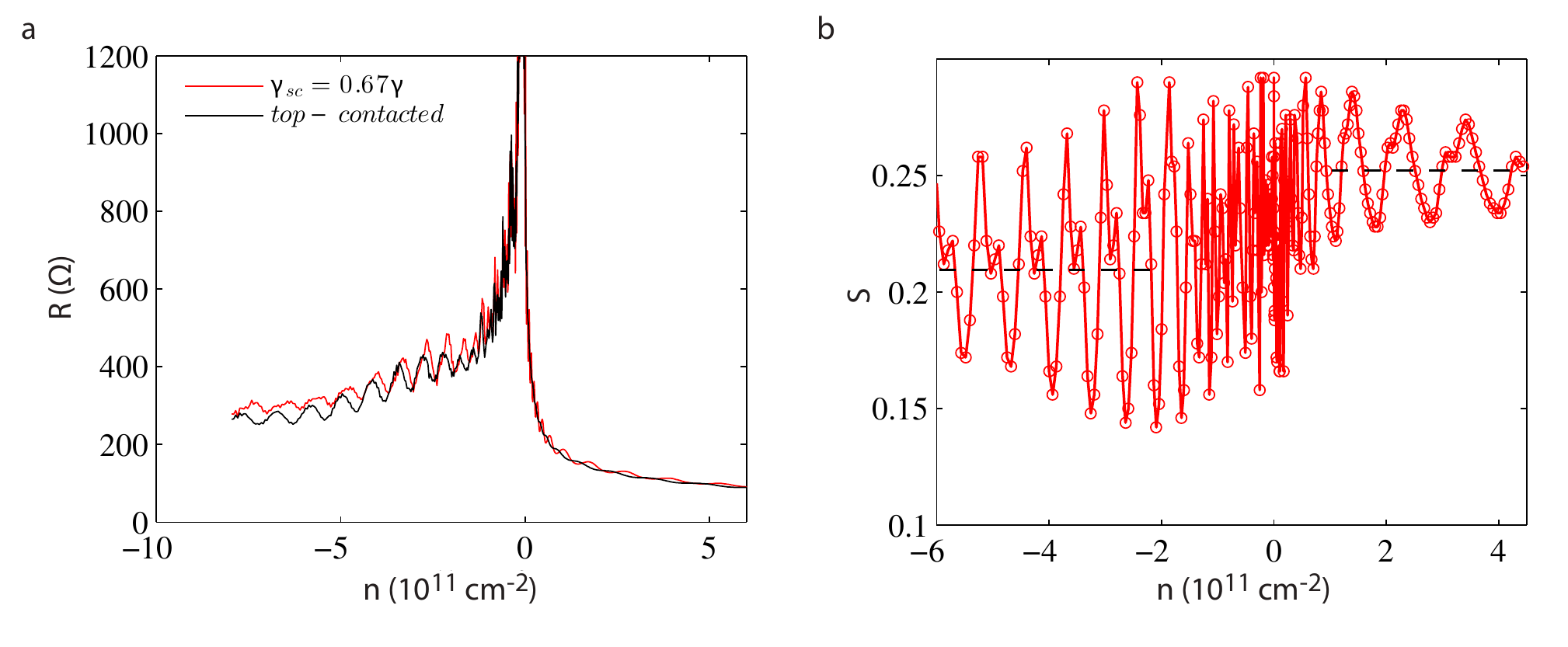}
	\caption{(a) Comparison of the calculated normal state resistance vs doping for top- and side-contacted 
		junctions, (b) Skewness vs doping calculated in the side-contact geometry. Dashed lines indicate
		the average skewness in the $n$ and $p$ doped regime.} 
	\label{fig:side-contact}
\end{figure}

We also performed  calculations using the side-contact geometry, which is shown in 
Figure~\ref{fig:contact-geometries}(b). 
This  contact geometry has recently been employed, e.g.,  in Reference~\cite{bouchiat_s} to model 
diffusive graphene JJs both in the short and in the long junction regime.  
The most important results of our calculations are shown in Figure \ref{fig:side-contact}. 
We have used the same doping profile $U(x)$ along the junction as in the top-contact geometry.
As it can be seen in  Figure \ref{fig:side-contact}(a), by choosing $\gamma_{sc}=0.67 \gamma$, 
the doping dependence of the   normal state resistance is qualitatively very similar for both models.
One can notice, however, that  the amplitude of the $R_n$ oscillations for  $nn'n$ doping is larger 
in the side-contact geometry. In the $npn$ regime the amplitude  of the  FP oscillations is 
somewhat different, but the oscillations are in the same phase, except for large $p$ doping. 

The skewness calculation for the side contact geometry is shown in Figure \ref{fig:side-contact}(b). 
We used the same $\Delta$ and  $\eta=0.17\Delta$ as for the corresponding calculation in the top-contact
geometry.
The result are qualitatively similar to those shown in  Figure \ref{fig:skewness-eta}  and 
Figure 4(c) of the main text. 
In particular, the average skewness is different in the $npn$ and $nn'n$ doping regime, but the obtained
$\bar{S}$ values are larger than the ones calculated in the top-contact geometry for $\eta=0.17\Delta$. 
However, the  amplitude of the skewness oscillations is larger in the side-contact 
geometry, especially for  $nn'n$ doping, where they are three times larger
than in Figure 4(c) of the main text. Such large oscillations are not present in the experimental 
data and for this reason
we find a better overall  agreement between the experiments the the calculations using the top-contact geometry.
Finally, we briefly note in the  vicinity of the CNP one can  see large oscillations in the skewness and 
therefore  both models fail to reproduce  the experimental results in this regime.


\end{document}